\documentclass[twocolumn,tighten]{aastex62}

\shorttitle{protoclusters at $z\sim4\mathrm{-}5$}
\shortauthors{Toshikawa et al.}

\begin{document}

\title{Discovery of Protoclusters at $z\sim3.7$ \& 4.9: Embedded in Primordial Superclusters}
 
\correspondingauthor{Jun Toshikawa}
\email{toshijun@icrr.u-tokyo.ac.jp}
\shorttitle{Discovery of protoclusters at $z\sim4\mathrm{-}5$}
\shortauthors{Toshikawa et al.}
\accepted{for publication in ApJ, December 2, 2019}

\author[0000-0001-5394-242X]{Jun Toshikawa}
\affil{Institute for Cosmic Ray Research, The University of Tokyo, Kashiwa, Chiba 277-8582, Japan.}
\affil{Department of Physics, University of Bath, Claverton Down, Bath, BA2 7AY, UK.}
\author{Matthew A. Malkan}
\affil{Department of Physics and Astronomy, University of California, Los Angeles, CA 90095-1547.}
\author{Nobunari Kashikawa}
\affil{Department of Astronomy, University of Tokyo, Hongo, Tokyo 113-0033, Japan.}
\affil{Optical and Infrared Astronomy Division, National Astronomical Observatory, Mitaka, Tokyo 181-8588, Japan.}
\author{Roderik Overzier}
\affil{Observat\'{o}rio Nacional, Rua Jos\'{e} Cristino, 77. CEP 20921-400, S\~{a}o Crist\'{o}v\~{a}o, Rio de Janeiro, RJ, Brazil}
\affil{Institute of Astronomy, Geophysics and Atmospheric Sciences, Department of Astronomy, University of S\~{a}o Paulo, Sao Paulo, SP 05508-090, Brazil}
\author{Hisakazu Uchiyama}
\affil{Department of Astronomy, School of Science, Graduate University for Advanced Studies, Mitaka, Tokyo 181-8588, Japan.}
\author{Kazuaki Ota}
\affil{Kyoto University Research Administration Office, Yoshida-Honmachi, Sakyo-ku, Kyoto 606-8501 Japan}
\author{Shogo Ishikawa}
\affil{Center for Computational Astrophysics, National Astronomical Observatory, Mitaka, Tokyo 181-8588, Japan.}
\author{Kei Ito}
\affil{Department of Astronomy, School of Science, Graduate University for Advanced Studies, Mitaka, Tokyo 181-8588, Japan.}

\begin{abstract}
We have carried out follow-up spectroscopy on three overdense regions of $g$- and $r$-dropout galaxies in the
Canada-France-Hawaii Telescope Legacy Survey Deep Fields, finding two new protoclusters at $z=4.898$, 3.721 and a
possible protocluster at $z=3.834$.
The $z=3.721$ protocluster overlaps with a previously identified protocluster at $z=3.675$.
The redshift separation between these two protoclusters is $\Delta z=0.05$, which is slightly larger than the size
of typical protoclusters.
Therefore, if they are not the progenitors of a $>10^{15}\,\mathrm{M_\sun}$ halo, they would grow into
closely-located independent halos like a supercluster.
The other protocluster at $z=4.898$ is also surrounded by smaller galaxy groups.
These systems including protoclusters and neighboring groups are regarded as the early phase of
superclusters.
We quantify the spatial distribution of member galaxies of the protoclusters at $z=3.675$ and 3.721 by fitting
triaxial ellipsoids, finding a tentative difference: one has a pancake-like shape while the other is
filamentary.
This could indicate that these two protoclusters are in different stages of formation.
We investigate the relation between redshift and the velocity dispersion of protoclusters, including other
protoclusters from the literature, in order to compare their dynamical states.
Although there is no significant systematic trend in the velocity dispersions of protoclusters with redshift,
the distribution is skewed to higher velocity dispersion over the redshift range of $z=2\mathrm{-}6$.
This could be interpreted as two phases of cluster formation, one dominated by the steady accretion of galaxies,
and the other by the merging between group-size halos, perhaps depending on the surrounding large-scale
environments.
\end{abstract}


\section{Introduction \label{sec:intro}}
The universe starts with a nearly uniform distribution of dark matter, and gravity gradually attracts matter to
local density peaks to make virialized halos.
The inhomogeneity of mass density keeps increasing; especially, mass density grows non-linearly in higher-density
regions \citep[cf.,][]{peebles80}.
Galaxies are formed according to the distribution of dark matter halos.
As dark matter halos become more massive by merging with surrounding halos, the contrast of the number density of
galaxies will also be higher.
In the local universe, as the result of hierarchical structure formation across cosmic time, we can see various
regions such as galaxy clusters, groups, filaments, or voids, which compose the large-scale structure of the
universe or cosmic web \citep[e.g.,][]{lapparent86,geller89,alpaslan14,libeskind18}.
Galaxy clusters are usually located at the knots of the cosmic web; especially, massive clusters tend to be
surrounded by other clusters or groups, known as ``superclusters,'' which are overdense regions in a few tens of
Mpc scale.
About half of local clusters are found to reside in superclusters based on the Abell's cluster catalog or X-ray
survey \citep{bahcall84,chon13}.
Therefore, galaxy clusters are key components of the large-scale structure of the universe, where mass density is
drastically increased over the initial small fluctuation.

In addition, galaxy clusters are good laboratories to understand environmental effects on galaxy evolution.
In the local universe, it is well known that galaxy properties in higher-density regions are significantly
different from those in lower-density regions: red, massive elliptical galaxies tend to reside in galaxy clusters,
and cluster galaxies make a tight sequence on a color-magnitude diagram
\citep[e.g.,][]{dressler80,lewis02,kauffmann04,bamford09}.
The stellar populations of these galaxies imply that they are generally formed at higher redshifts than field
counterparts, and experience a short and intense star-formation activity like star-burst phase early in their
formation history \citep{thomas05,raichoor11,gu18}.
Furthermore, the physical properties of brightest cluster galaxies depend on the internal structure or dynamical
state of their host clusters at $z\lesssim0.5$ \citep{wen13}, and, in the local universe, the morphology of
superclusters correlates with e.g., stellar mass and star-formation rate (SFR) of member galaxies
\citep{einasto14}.
Although in the local universe, we can see the differences of galaxy properties, which are attributed to
environmental effects, it is still unclear when and how galaxies are affected by surrounding environments.
In parallel with galaxy evolution in high-density environments, the large-scale structure itself is developing
over cosmic time.
The redshift evolution of both galaxies and large-scale structure is intricately connected due to anisotropic
galaxy/mass assembly along to filaments \citep{kraljic18}.
This complexity would prominently appear in galaxy clusters, as they are at the knots of the cosmic web.
Thus the existence of environmental effects on galaxy evolution are clearly confirmed by the studies of local
galaxy clusters.
However, in order to reveal the physical mechanisms of environmental effects over the long history of cluster
formation, we need to directly observe the early stage of cluster formation, which would allow us to understand
the physical properties on their way to mature galaxy clusters. 
The progenitors of galaxies clusters at high redshifts, or protoclusters, are good laboratories for investigating
the relation between galaxy evolution and cluster formation \citep{overzier16}.

Galaxy clusters having extended X-ray emission are found up to $z=2.5$ \citep{gobat13,wang16}, and quiescent
galaxies reside in some clusters at $z\sim2$ \citep{newman14,strazzullo18}.
Beyond $z\sim2$, young star-forming galaxies, such as Lyman break galaxies (LBGs) and Ly$\alpha$ emitters (LAEs),
tend to be a dominant galaxy population even in high-density regions \citep{kuiper10,spitler12,contini16}
though some quiescent galaxies are also clearly found in protoclusters at $z\sim2\mathrm{-}3$
\citep{kodama07,kubo13,shi19a}.
Thus, protoclusters are found to harbor the wide range of galaxy populations. 
\citet{shimakawa18} showed that protocluster galaxies at $z\sim2\mathrm{-}4$ are more actively forming stars than
in fields, and \citet{forrest17} found that extreme [O{\sc iii}]$+$H$\beta$ emission line galaxies are clustered
in an overdense region.
Similarly, dusty star-burst galaxies identified by submillimeter imaging are frequently discovered in
protoclusters \citep{casey16,umehata17,miller18,zeballos18}.
However, \citet{tran17} reported that H$\alpha$ emitters exhibit similar stellar growth regardless of environments.
It is still unclear what causes the diversity of protocluster properties, such as star-forming activity. 
Some studies imply that there is a large amount of cold gas around protoclusters \citep{cucciati14,lemaux18}.
Such cold gas could enhance the star-formation of protocluster galaxies, or ignite star-burst if it falls into the
core of a halo as cold stream.
Even if the total amount of cold gas around protoclusters is the same, how much star-formation is enhanced can
largely vary because the accretion rate of such cold gas is dependent on surrounding large-scale structures such
as the number of filaments connected to knots \citep{dekel09,aragon10,liao19}.
Although there are other possible physical mechanisms, investigating protoclusters from the viewpoints of the
large-scale structure is one of the approaches to reveal galaxy evolution in high-density environments.

However, the rarity of protoclusters at high redshifts makes it difficult to conduct a systematic study.
So far, the number of known protoclusters is only a few tens at $z\gtrsim2$ (only $\sim10$ at $z\gtrsim4$).
To find such rare objects, many studies have used radio galaxies (RGs) or quasars (QSOs) as the signpost of
overdense regions \citep[e.g.,][]{venemans07,wylezalek13} because such galaxies are expected to be located in
massive dark matter halos \citep[e.g.,][]{shen07,orsi16}.
However, the relation between these objects and environment is still under debate: \citet{noirot18} confirmed
protoclusters around RGs at $1.4<z<2.8$ while \citet{uchiyama18} found that there is no correlation between QSOs
and environments at $z\sim4$.
The fraction of active galactic nuclei (AGNs) is different among protoclusters
\citep{lehmer13,krishnan17,macuga19}.
Thus, the method of using signposts has the potential to pick up only a subset of protoclusters.
Complementary protocluster searches based on blank surveys without such signposts of protoclusters have been
extensively performed recently.
For example, the spectroscopic survey of the VIMOS Ultra-Deep Survey \citep{fevre15} has found many protoclusters
at $z\sim3\mathrm{-}5$ by the direct investigation of spatial and redshift clustering of galaxies
\citep{lemaux14,cucciati14,lemaux18,cucciati18}; the wide-field imaging survey of the Hyper SuprimeCam
\citep{aihara18} has made a systematic sample of protocluster candidates up to $z\sim6.6$ based on the projected
overdensity of LBGs and LAEs \citep{toshikawa18,higuchi19}.
It should be noted that even blank searches would identify only another subset of protoclusters.
Photometric surveys require a certain selection of galaxy population; on the other hand, spectroscopic surveys
tend to observe brighter targets compared with photometric surveys though most of them observe down to well below
the characteristic luminosity at a given redshift.
From these various searches, the number of known protoclusters are gradually increasing \citep{badescu17,oteo18},
which enables us to see a large variety of protoclusters (e.g., overdensity, size, galaxy population, and physical
properties of member galaxies).

\citet[][hereafter T16]{toshikawa16} have also carried out a blank protocluster search in the $4\,\mathrm{deg^2}$
area of the Canada-France-Hawaii Telescope Legacy Survey (CFHTLS) Deep Fields \citep{gwyn12}, and identified 21
protocluster candidates at $z\sim3\mathrm{-}6$, which are defined as $>4\sigma$ significance overdense regions of
$u$-, $g$-, $r$-, or $i$-dropout galaxies.
By comparison with theoretical model \citep{henriques12}, 76\% of candidates are expected to be in real
protoclusters.
Following this search for protocluster candidates, a follow-up spectroscopic observation is conducted on several
of them.
Three among four spectroscopically observed candidates are confirmed as genuine protoclusters at $z=3.13$, 3.24,
and 3.67 with more than five members spectroscopically confirmed.
Although we have made follow-up spectroscopy for only four candidates at $z\sim3\mathrm{-}4$, this success rate is
consistent with theoretical expectation.
In other two candidates, close galaxy pairs are found at $z=4.89$ and 5.75, suggesting the existence of
protoclusters, though more complete follow-up spectroscopy is required to reach a conclusion.
In this study, we will extend follow-up spectroscopy for the protocluster candidates found by T16 to discover more
protoclusters and make a close investigation into each protocluster.
We observe three overdense regions using KeckII/DEIMOS \citep{faber03}: the one includes a plausible protocluster
candidate at $z=4.89$ because a close galaxy pair was found; the second is not observed by the previous follow-up
spectroscopy of T16; the last contains a known protocluster at $z=3.67$ in order to make a more detailed
investigation by increasing the number of the identification of member galaxies.
The wide field-of-view (FoV) of KeckII/DEIMOS allows us to discuss the spatial and redshift distribution of
galaxies from the viewpoint of large-scale structure.
Here we present the results of our follow-up spectroscopy, including newly confirmed protoclusters.

The structure of this paper is as follows.
Section \ref{sec:obs} describes new observations and the details of targets.
In Section \ref{sec:res}, the results of follow-up spectroscopy are shown, and the probability of the existence of
protoclusters is estimated.
We discuss the primordial large-scale structure and the internal structures of protoclusters in Section
\ref{sec:disc}.
The conclusions are provided in Section \ref{sec:conc}.
We assume the following cosmological parameters: $\Omega_\mathrm{M}=0.3, \Omega_\Lambda=0.7,
\mathrm{H}_0=70 \mathrm{\,km\,s^{-1}\,Mpc^{-1}}$, and magnitudes are given in the AB system.

\section{Observations \label{sec:obs}}
\subsection{Targets}
We have obtained follow-up spectroscopy on three overdense regions in the CFHTLS Deep Fields, which are identified
by T16.
Here is a brief description of the imaging dataset and protocluster search in T16.
The CFHTLS Deep Fields consist of four separate fields, and, in each field, five optical broad-band datasets are
available over $\sim1\,\mathrm{deg^2}$.
The depth is almost uniform between field to field, and the $3\sigma$ limiting magnitudes are $\sim28.1$, 28.3,
27.8, 27.3, and $26.4\,\mathrm{mag}$ at $u$-, $g$-, $r$-, $i$-, and $z$-bands, respectively.
This corresponds to about $M^*_\mathrm{UV}+2$ at $z\sim4\mathrm{-}5$ \citep[where $M^*_\mathrm{UV}$ is the
characteristic magnitude of the Schechter function;][]{bouwens07,burg10}.
From these wide and deep fields, we have selected  $u$-, $g$-, $r$-, and $i$-dropout galaxies by the standard
Lyman break technique over $\sim4\,\mathrm{deg^2}$ area.
Local surface galaxy number density is calculated by counting dropout galaxies within an aperture of
$0.75\,(1.0)\,\mathrm{Mpc}$ radius in physical scale for $u$-, $g$-, and $r$-dropout ($i$-dropout) galaxies
because about 65\% of mass of the progenitors of $1\mathrm{-}3\times10^{14}\,M_\sun$ halos is enclosed in this
radius according to theoretical predictions \citep{chiang13}. 
Although larger apertures can include protocluster galaxies more completely, the excess of number density by a
protocluster will weaken due to the contamination of fore/background galaxies due to the large redshift
uncertainty of Lyman break technique ($\Delta z\sim1$).
The apertures are distributed over the whole CFHTLS Deep Fields, and protocluster candidates are defined as
regions where the number density excess from the average is $>4\sigma$ significance.
Comparing with a theoretical model \citep{henriques12}, 76\% of $>4\sigma$ overdense regions are expected to grow
into galaxy clusters with the halo mass of $>10^{14}\,\mathrm{M_\sun}$ (refer to T16 for the details).
Due to the large redshift uncertainty of Lyman break selection, the completeness of our protocluster search is
very small ($\sim10\%$).
In particular, the progenitors of smaller galaxy clusters would be largely affected.
Thus, it should be noted that our search by using dropout galaxies preferentially identifies more massive
protoclusters.

\begin{deluxetable*}{ccccccc}
\tablecaption{Overview of the targets of the follow-up spectroscopy  \label{tab:target}}
\tablewidth{0pt}
\tablehead{Name & R.A. (J2000) & Decl. (J2000) & Population & Overdensity\tablenotemark{a} &
    $N_\mathrm{LBG}$\tablenotemark{b} & T16\tablenotemark{c}}
\startdata
D1RD01 & 02:24:45.3 & $-$04:55:56.5 & $r$-dropout & $4.4\sigma$ & 40 & Yes (15) \\
D1GD02 & 02:25:56.2 & $-$04:48:30.4 & $g$-dropout & $4.2\sigma$ & 153 & No \\
D4GD01 & 22:16:47.3 & $-$17:16:52.7 & $g$-dropout & $4.3\sigma$ & 153 & Yes (60) \\
\enddata
\tablenotetext{a}{Overdensity at the peak.}
\tablenotetext{b}{Number of dropout galaxies within $3\,\mathrm{arcmin}$ radius from the overdensity peak.
    Noted that DEIMOS can observe more galaxies since its FoV is wider than $6\,\mathrm{arcmin}$.}
\tablenotetext{c}{The overdense regions observed by follow-up spectroscopy in T16 are marked as ``Yes.''
    The number of spectroscopically observed galaxies in T16, which are located within $3\,\mathrm{arcmin}$
    radius from its overdensity peak, is described in the parenthesis.}
\vspace{-10mm}
\end{deluxetable*}

By this criterion, 21 protocluster candidates are identified from $z\sim3$ to $z\sim6$.
Eight (two at each redshift) of them were observed by follow-up spectroscopy in T16.
Based on the theoretical model, we have evaluated the spatial distribution of protocluster members, which will
merge into the same halo at $z=0$ (see Figure 8 in T16), and a typical redshift size of protoclusters is found to
be $\Delta z\lesssim0.03$.
Then, three of the targeted eight candidates show strong redshift clustering within this redshift range, and we
have investigated whether these observed concentrations can coincidentally be reproduced from a random galaxy
distribution drawn from the redshift selection function of dropout galaxies.
As a result, they cannot be reproduced from a random distribution $>99\%$ of the time.
Therefore, these three candidates are confirmed to be real protoclusters at $z=3.13$, 3.24, and 3.67.
In the same manner, since the redshift distribution in one overdense region of $g$-dropout galaxies is consistent
with a random distribution probability of 21\%, it is not regarded as a confirmed protocluster.
In the other four candidates of $r$- and $i$-dropout galaxies, we cannot conclude whether they are real
protoclusters or not because of the insufficient follow-up observations for higher-redshift candidates.
The interested reader should refer to T16 for more details.

In this study, we focus on protocluster candidates at $z\gtrsim4$ because the number of known protoclusters is
particularly small at these redshifts.
Also, it would be observationally difficult to confirm protoclusters at $z\sim6$ ($i$-dropouts) due to the shallow
images of $z$-band in the CFHTLS Deep Fields.
Therefore, we have made follow-up spectroscopic observations of the three overdense regions of $r$- and $g$-dropout
galaxies in the D1 and D4 fields, which are termed ``D1RD01'', ``D1GD02'', and ``D4GD01'' in T16 respectively.
Table \ref{tab:target} shows the basic information of these three target regions (e.g., R.A., Decl., or
overdensity).
The overdense regions of D1RD01 and D4GD01 were already observed by follow-up spectroscopy in T16, while follow-up
spectroscopy is for the first time performed for the overdense region of D1GD02 in this study.
In the two overdense regions which are spectroscopically observed in T16, the slits of the previous spectroscopic
observation are allocated to less than half of the dropout galaxies.
Thus, even for the previously observed overdense regions, further follow-up spectroscopy is necessary in order to
make a closer investigation and to draw firm conclusions about the possible protocluster.
We briefly summarize information regarding the follow-up spectroscopy for these three target regions below, and
give further details in Section 4 of T16.
\begin{description}
\item[D1RD01]
We have identified the redshifts of only six $r$-dropout galaxies in total.
Although two galaxies out of six are clustered in redshift as well as spatial space, this is too small a number
to conclude that it is a protocluster.
The high overdensity of the projected number of galaxies can be attributed to the chance alignment of some small
groups along the line of sight, instead of a single massive structure like a protocluster.
\item[D1GD02]
This region is not the target of the previous follow-up spectroscopy of T16.
\item[D4GD01]
The overdense region of D4GD01 is already confirmed to include a protocluster at $z=3.67$ composed of eleven
member galaxies at least.
An AGN is also found in this region, but it would be hard to regard it as a member of the protocluster because
the AGN is located far behind the protocluster (line-of-sight separation between the AGN and the protocluster is
$\Delta z=0.05$ or $\sim8\,\mathrm{Mpc}$ in the physical scale at $z=3.7$).
\end{description}

\subsection{Follow-up Spectroscopy}
\begin{deluxetable*}{ccccccc}
\tablecaption{Overview of our spectroscopic observations\tablenotemark{*} \label{tab:obs}}
\tablewidth{0pt}
\tablehead{Target & Date & Grism & $t_\mathrm{exp}$ (min) & Seeing & $N_\mathrm{obs}$\tablenotemark{a} &
    $N_\mathrm{det}$\tablenotemark{b}}
\startdata
D1RD01 & 2015 Sep 14 & 830G & 175 & $0\farcs8$ & 47 (4) & 15 \\
 & 2016 Sep 9 \& 10 & 900ZD & 313 & $0\farcs9$ & 38 (20) & 8 \\
D1GD02 & 2015 Sep 15 & 900ZD & 157 & $0\farcs8$ & 101 & 21 \\
 & 2016 Oct 28 & 900ZD & 270 & $0\farcs7$ & 85 (2) & 29 \\
D4GD01 & 2016 Oct 28 & 900ZD & 200 & $0\farcs7$ & 90 (9) & 10 \\
\enddata
\tablenotetext{*}{Each row shows the information on one mask.}
\tablenotetext{a}{Number of observed galaxies.
    The number of galaxies observed by previous observations is given in the parenthesis.
    A DEIMOS mask can typically contain $\sim90\mathrm{-}100$ slits; thus, we have also observed dropout galaxies
    at other redshifts as mask fillers.
    For example, the masks targeting $r$-dropout overdense regions include $g$-dropout galaxies as well.
    Since such mask fillers do not belong to the candidates of protoclusters, they are not used in this study.}
\tablenotetext{b}{Number of spectroscopically detected galaxies.}
\vspace{-10mm}
\end{deluxetable*}

Our spectroscopic observations in the two overdense regions were conducted as part of the Keck Observatory
programs of U033D in 2015 and U130D in 2016.
We used KeckII/DEIMOS with Multi-Object Spectroscopy (MOS) mode.
The slits have a length of $4.0\,\mathrm{arcsec}$ at minimum and a width of $1.0\,\mathrm{arcsec}$.
We use the gratings of 830G and 900ZD, which have high efficiency at the wavelengths of
$\sim5000\mathrm{-}8000\,\mathrm{\AA}$ corresponding to the wavelength of the redshifted Ly$\alpha$ emission line
of $g$- or $r$-dropout galaxies.
The spectral resolution of this configuration ($2.1\mathrm{-}2.5\,\mathrm{\AA}$) is high enough to resolve the
[O{\sc ii}] doublet (the wavelength separation is $3.9\mathrm{-}5.6\,\mathrm{\AA}$ in the observed-frame).
The wide spectral range of DEIMOS enables us to fully cover the expected wavelength range of Ly$\alpha$ emissions
from $g$- or $r$-dropout galaxies, or to detect H$\alpha$, H$\beta$, and [O{\sc iii}] emission lines simultaneously
if they are from contaminating low-redshift galaxies.
Therefore we can distinguish a single Ly$\alpha$ emission line from these other possible low-redshift contaminants.
Furthermore, we calculate weighted skewness, $S_w$, which is a good indicator of asymmetry, to distinguish
Ly$\alpha$ emission line from [O{\sc ii}] doublet \citep{kashikawa06}.
In case that doublet is detected as a single line due to low spectral resolution, it should not show a large
skewness.
Section 4.2 of T16 gives more details about line contaminations.

We have used five masks in total: two masks for the D1RD01, two for the D1GD02, and one for the D4GD01 region.
The details of configuration and sky condition for each mask observation are summarized in Table \ref{tab:obs}.
We put a higher priority for allocating slits to galaxies which were not observed by the previous follow-up
spectroscopy in T16.
Our masks also include observed but unconfirmed galaxies in order to detect their possible faint Ly$\alpha$
emission.
In particular, the mask for the overdense region of D1RD01 used in Sep., 2016 has many duplicated targets.
The pipeline of spec2d\footnote{The data-reduction pipeline was developed at the University of California,
Berkeley, with support from National Science Foundation grant AST 00-71048.} is used to reduce the data taken by
DEIMOS \citep{cooper12,newman13}.
The pipeline involves dividing the raw images into individual slits, flat-fielding, calculating wavelength
solution, subtracting sky background, and combining separate exposures into one spectrum with cosmic-ray removal.
In addition to the science targets, slits in each mask are allocated for bright stars ($\sim20\,\mathrm{mag}$) to
monitor the time variations of seeing size or atmospheric transmission between exposures.
The differences of seeing size and transparency between exposures are found to be $\lesssim0\farcs1$ and
$\lesssim10\%$, respectively.
Table \ref{tab:obs} also shows seeing sizes and total exposure times, and the integration time for each individual
exposure is typically 20 minutes.

\begin{figure*}
\epsscale{1.0}
\plotone{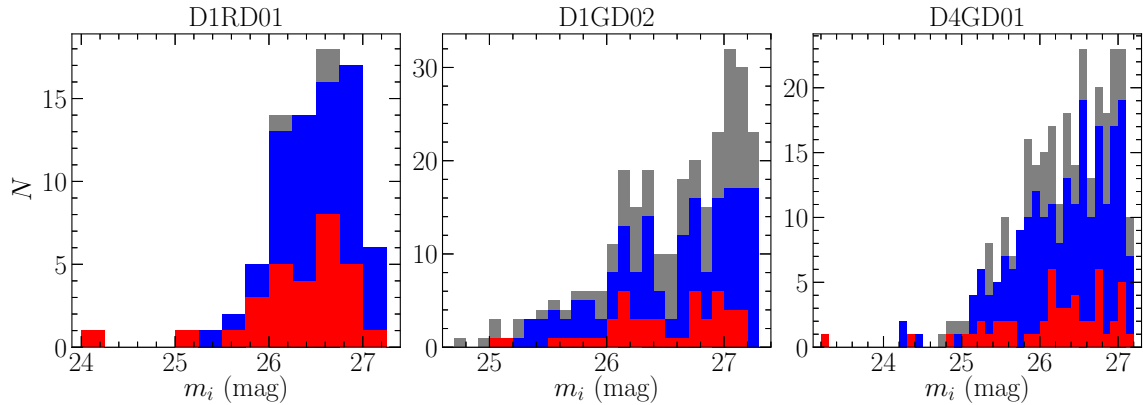}
\caption{$i$-band magnitude distributions of dropout (gray), spectroscopically-observed (blue), and
    Ly$\alpha$-detected galaxies (red) within $3\,\mathrm{arcmin}$ radius of the overdensity peaks of the D1RD01
    (left), D1GD02 (middle), and D4GD01 regions (right).}
\label{fig:spmag}
\end{figure*}

Based on one-dimensional spectra produced by the pipeline, we have made the crude identification of possible
emission lines with the criterion of three connected pixels having signal-to-noise ratio ($S/N$) more than 1.0 per
pixel.
The fake emission lines caused by sky residuals and ghosts can be removed by visual inspection on two-dimensional
spectra.
Because the predicted position from the mask design could be shifted by up to a few pixel, one-dimensional spectra
are manually extracted again so that we can correctly trace object positions.
Then, we have estimated the $S/N$ of emission lines by integrating all pixels in a line profile, and insignificant
lines with $S/N<3$ are removed from the sample of detected emission lines.
For flux calibration, we have observed the spectroscopic standard stars of Feige15 or BD+28d4211 with a
$1.0\,\mathrm{arcsec}$ long slit and the same grating as science targets each night.
The standard stars are reduced in the same way as science targets, and the slit loss is corrected based on the
ratio between slit width and seeing size, in which its light profile is assumed to be a Gaussian function whose
width is the seeing size.
The sensitivity as a function of wavelength is estimated by the IRAF task {\sf standard} and {\sf sensfunc} with
the correction of airmass and extinction, and applied to science targets by the IRAF task {\sf fluxcalib}.

In this study, we have allocated 361 slits for dropout galaxies in the overdense regions of $r$- and $g$-dropout
galaxies, and 83 galaxies are newly confirmed by detecting their Ly$\alpha$ emission lines (Table \ref{tab:obs}).
Although our spectroscopy may detect continuum only from bright galaxies ($\lesssim24\,\mathrm{mag}$), its $S/N$
is not high enough to identify absorption lines, and we cannot precisely determine redshifts by the Lyman break.
Therefore, it should be noted that we can spectroscopically confirm only dropout galaxies having a Ly$\alpha$
emission.
Combining with the previous work of T16, the total numbers of spectroscopically-observed(confirmed) galaxies are
76(29), 184(50), and 224(52) in the overdense regions of D1RD01, D1GD02, and D4GD01, respectively.
Figure \ref{fig:spmag} shows the magnitude distributions of dropout and spectroscopically-observed/confirmed
galaxies in the three target regions.
The fractions of spectroscopically-observed galaxies among dropout galaxies located within $3\,\mathrm{arcmin}$
radius from the overdensity peak are 93\%($=37/40$), 32\%($=49/153$), and 53\%($=81/153$) for the D1RD01, D1GD02,
and D4GD01 regions, respectively.
Based on the Kolmogorov-Smirnov (KS) test, the magnitude distributions of spectroscopically-observed/confirmed
galaxies are consistent with that of dropout galaxies located in the overdense regions (the $p$-values of the KS
test are $p_\mathrm{KS}>0.5$ for any combinations).
There is no clear contamination of low-redshift galaxies in our follow-up spectroscopy.
However, since the possible contaminants of the color selection of dropout galaxies are mainly dwarf stars or
quiescent galaxies rather than H$\alpha$, [O{\sc ii}], or [O{\sc iii}] emitters, we need continuum detections with
high $S/N$ in order to spectroscopically confirm them.
Although faint low-redshift galaxies or dwarf stars could contaminate the sample of unconfirmed dropout galaxies,
the contamination rate expected by the color selection of dropout galaxies may be up to $\sim5\%$ at worst (refer
T16 for the details).

We use only confirmed dropout galaxies for the following analysis.
The observed properties, such as redshift, Ly$\alpha$ luminosity ($L_\mathrm{Ly\alpha}$), UV absolute magnitude
($M_\mathrm{UV}$), and rest-frame Ly$\alpha$ equivalent width ($EW_0$), of newly confirmed galaxies are described
in Table \ref{tab:spec}, and their one- and two-dimensional spectra are shown in Figure \ref{fig:spec}.
Their IDs are continued from T16; thus, ID=1-6 in the D1RD01 region and ID=1-42 in the D4GD01 region are also
shown in Table 4 and Figure 9 of T16.
The redshifts are derived by the peak wavelength of the Ly$\alpha$ emission line, assuming the rest wavelength of
Ly$\alpha$ to be 1215.6{\AA}.
These measurements could be overestimated if there is a galactic outflow.
When emission lines are located near strong sky lines, the position of the peak could be shifted.
These effects of sky lines and the wavelength resolution are taken into account when estimating the uncertainty.
The observed line flux, $f_\mathrm{Ly\alpha}$, corresponds to the total amount of the flux within the line profile.
The slit loss is corrected based on the ratio of slit width and seeing size for each observation, and its
uncertainty is estimated from the combination of the line width and the noise level at wavelengths blueward of
Ly$\alpha$.
Since continuum flux is too faint to be detected in the observed spectra, $M_\mathrm{UV}$ is estimated from the
broadband photometry.
It is derived from $r$-band ($i$-band) magnitudes for $g$-dropout ($r$-dropout) galaxies after correcting the
contribution of the absorption of intergalactic medium (IGM) and the Ly$\alpha$ emission.
In this calculation, we assume flat UV spectra ($f_\lambda\propto\lambda^\beta$ where $\beta=-2$) and the IGM
transmission model of \citet{madau95}.
It should be noted that the shape of UV spectra can differ according to various galaxy properties \citep[e.g.,
dust, age, or metallicity;][]{bouwens12}, and IGM absorption also varies depending on the line of sight
\citep{thomas17}.
Although it is difficult to predict physical properties with these UV spectra, an observed broad-band magnitude
can be converted into pure continuum flux with the spectroscopic measurements of $f_\mathrm{Ly\alpha}$ and
redshift.
We have confirmed that $M_\mathrm{UV}$ changes $\sim5\%$ when UV slope or IGM transmission fluctuate between
$\pm1.0$ or $\pm15\%$ respectively.
This systematic error is smaller than the photometric error of our dataset.
In addition, $EW_0$ is estimated by combining $f_\mathrm{Ly\alpha}$ and $M_\mathrm{UV}$.

\startlongtable
\begin{deluxetable*}{cccccccccc}
\tabletypesize{\scriptsize}
\tablecaption{Observed properties of spectroscopically confirmed dropout galaxies. \label{tab:spec}}
\tablewidth{0pt}
\tablehead{
    \colhead{ID} & \colhead{R.A.} & \colhead{Decl.} & \colhead{$m_i$} & \colhead{Redshift} &
        \colhead{$M_\mathrm{UV}$} & \colhead{$f_\mathrm{Ly\alpha}$} & \colhead{$L_\mathrm{Ly\alpha}$} &
        \colhead{$EW_0$} & \colhead{$S_w$} \\
    \colhead{} & \colhead{(J2000)} & \colhead{(J2000)} & \colhead{(mag)} & \colhead{} & \colhead{(mag)} &
        \colhead{($10^{-18}\,\mathrm{erg\,s^{-1}\,cm^{-2}}$)} & \colhead{($10^{42}\,\mathrm{erg\,s^{-1}}$)} &
        \colhead{(\AA)} & \colhead{(\AA)}
}
\startdata
\multicolumn{10}{c}{D1RD01 (23 galaxies)} \\
7 & 02:24:33.40 & -04:57:58.4 & $26.92\pm0.09$ & $4.473^{+0.001}_{-0.001}$ & $-19.33\pm0.23$ & $1.90\pm0.30$ & $0.38\pm0.06$ & $8.00\pm2.25$ & $10.11\pm1.99$ \\
8 & 02:24:58.59 & -04:56:25.6 & $26.36\pm0.06$ & $4.602^{+0.001}_{-0.001}$ & $-19.93\pm0.14$ & $3.22\pm0.48$ & $0.69\pm0.10$ & $8.34\pm1.71$ & $4.15\pm0.80$ \\
9 & 02:25:24.73 & -04:53:10.1 & $25.13\pm0.02$ & $4.635^{+0.001}_{-0.001}$ & $-21.17\pm0.05$ & $6.39\pm0.89$ & $1.39\pm0.19$ & $5.38\pm0.79$ & $4.44\pm1.79$ \\
10 & 02:24:52.51 & -04:56:08.5 & $24.21\pm0.01$ & $4.648^{+0.001}_{-0.001}$ & $-22.10\pm0.02$ & $20.07\pm1.02$ & $4.40\pm0.22$ & $7.22\pm0.39$ & $9.68\pm0.56$ \\
11 & 02:25:20.08 & -04:52:54.8 & $26.91\pm0.09$ & $4.671^{+0.001}_{-0.001}$ & $-19.41\pm0.23$ & $1.06\pm0.30$ & $0.24\pm0.07$ & $4.62\pm1.69$ & $8.26\pm9.97$ \\
12 & 02:25:33.61 & -04:56:43.1 & $26.37\pm0.06$ & $4.724^{+0.001}_{-0.001}$ & $-19.96\pm0.15$ & $4.01\pm0.60$ & $0.91\pm0.14$ & $10.80\pm2.24$ & $4.15\pm1.55$ \\
13 & 02:24:31.90 & -04:55:46.6 & $26.00\pm0.04$ & $4.850^{+0.001}_{-0.001}$ & $-20.12\pm0.14$ & $21.90\pm1.45$ & $5.31\pm0.35$ & $53.77\pm8.25$ & $10.47\pm1.02$ \\
14 & 02:24:30.17 & -04:55:59.5 & $25.97\pm0.04$ & $4.851^{+0.001}_{-0.001}$ & $-20.24\pm0.13$ & $17.51\pm1.17$ & $4.24\pm0.28$ & $38.77\pm5.46$ & $9.84\pm0.99$ \\
15 & 02:24:52.95 & -04:57:56.2 & $25.96\pm0.04$ & $4.889^{+0.001}_{-0.001}$ & $-20.31\pm0.12$ & $15.23\pm0.56$ & $3.76\pm0.14$ & $31.99\pm4.03$ & $9.31\pm0.63$ \\
16 & 02:25:32.22 & -04:55:40.0 & $26.65\pm0.07$ & $4.892^{+0.001}_{-0.001}$ & $-19.87\pm0.19$ & $1.12\pm0.27$ & $0.28\pm0.07$ & $3.56\pm1.08$ & $3.06\pm2.01$ \\
17 & 02:24:47.88 & -04:54:28.9 & $26.52\pm0.07$ & $4.898^{+0.001}_{-0.001}$ & $-19.71\pm0.21$ & $11.60\pm1.03$ & $2.88\pm0.26$ & $42.64\pm10.04$ & $7.27\pm0.97$ \\
18 & 02:24:51.79 & -04:54:56.7 & $26.18\pm0.05$ & $4.907^{+0.001}_{-0.001}$ & $-20.33\pm0.13$ & $2.14\pm0.44$ & $0.53\pm0.11$ & $4.46\pm1.07$ & $4.11\pm2.44$ \\
19 & 02:24:32.96 & -04:55:05.0 & $26.51\pm0.06$ & $4.914^{+0.001}_{-0.001}$ & $-19.79\pm0.20$ & $9.96\pm0.87$ & $2.49\pm0.22$ & $34.43\pm7.65$ & $9.55\pm1.19$ \\
20 & 02:25:16.85 & -04:57:01.3 & $25.95\pm0.04$ & $4.943^{+0.001}_{-0.001}$ & $-20.62\pm0.10$ & $2.46\pm0.36$ & $0.62\pm0.09$ & $4.00\pm0.71$ & $4.59\pm1.22$ \\
21 & 02:25:20.12 & -04:53:10.0 & $26.55\pm0.07$ & $4.949^{+0.001}_{-0.001}$ & $-19.94\pm0.18$ & $4.40\pm0.70$ & $1.12\pm0.18$ & $13.37\pm3.26$ & $2.22\pm1.07$ \\
22 & 02:25:26.32 & -04:54:32.8 & $26.50\pm0.06$ & $4.958^{+0.002}_{-0.001}$ & $-19.54\pm0.27$ & $20.08\pm0.85$ & $5.13\pm0.22$ & $88.61\pm24.87$ & $5.38\pm0.32$ \\
23 & 02:24:42.76 & -04:55:45.3 & $26.56\pm0.07$ & $5.056^{+0.001}_{-0.001}$ & $-20.06\pm0.20$ & $6.80\pm0.70$ & $1.82\pm0.19$ & $19.49\pm4.35$ & $6.61\pm1.21$ \\
24 & 02:25:16.35 & -04:55:04.9 & $25.70\pm0.03$ & $5.090^{+0.001}_{-0.001}$ & $-21.01\pm0.09$ & $12.58\pm1.31$ & $3.41\pm0.35$ & $15.26\pm2.07$ & $6.29\pm0.88$ \\
25 & 02:25:32.46 & -04:54:37.8 & $26.51\pm0.06$ & $5.107^{+0.001}_{-0.001}$ & $-20.25\pm0.18$ & $4.08\pm0.81$ & $1.11\pm0.22$ & $10.04\pm2.67$ & $5.92\pm2.33$ \\
26 & 02:25:18.42 & -04:55:53.6 & $26.61\pm0.07$ & $5.173^{+0.001}_{-0.001}$ & $-20.23\pm0.20$ & $5.39\pm0.47$ & $1.52\pm0.13$ & $13.97\pm3.03$ & $10.67\pm1.41$ \\
27 & 02:25:17.94 & -04:57:24.9 & $26.86\pm0.09$ & $5.402^{+0.001}_{-0.001}$ & $-19.83\pm0.38$ & $19.41\pm1.23$ & $6.05\pm0.38$ & $80.58\pm34.20$ & $3.96\pm0.88$ \\
28 & 02:25:24.23 & -04:54:25.4 & $26.75\pm0.08$ & $5.402^{+0.001}_{-0.001}$ & $-20.26\pm0.27$ & $12.94\pm0.78$ & $4.04\pm0.24$ & $36.05\pm10.33$ & $4.32\pm0.70$ \\
29 & 02:25:33.51 & -04:54:16.7 & $27.07\pm0.11$ & $5.470^{+0.001}_{-0.001}$ & $-20.19\pm0.32$ & $7.26\pm0.54$ & $2.33\pm0.17$ & $22.25\pm7.70$ & $8.17\pm0.89$ \\
\hline
\multicolumn{10}{c}{D1GD02 (50 galaxies)} \\
1 & 02:26:02.53 & -04:49:03.0 & $26.20\pm0.05$ & $3.636^{+0.001}_{-0.001}$ & $-19.70\pm0.08$ & $4.42\pm0.72$ & $0.54\pm0.09$ & $8.11\pm1.47$ & $3.35\pm1.14$ \\
2 & 02:25:59.84 & -04:50:37.0 & $25.51\pm0.03$ & $3.638^{+0.001}_{-0.001}$ & $-20.54\pm0.04$ & $84.45\pm1.35$ & $10.36\pm0.17$ & $71.69\pm2.88$ & $8.18\pm0.49$ \\
3 & 02:25:11.74 & -04:47:47.8 & $25.80\pm0.03$ & $3.674^{+0.001}_{-0.001}$ & $-20.02\pm0.06$ & $3.25\pm0.46$ & $0.41\pm0.06$ & $4.56\pm0.70$ & $5.56\pm0.97$ \\
4 & 02:25:18.03 & -04:49:13.7 & $27.01\pm0.10$ & $3.679^{+0.001}_{-0.001}$ & $-18.43\pm0.26$ & $3.74\pm0.69$ & $0.47\pm0.09$ & $22.71\pm7.36$ & $4.24\pm1.82$ \\
5 & 02:25:49.68 & -04:48:13.9 & $26.87\pm0.09$ & $3.681^{+0.001}_{-0.001}$ & $-19.08\pm0.15$ & $3.97\pm0.81$ & $0.50\pm0.10$ & $13.30\pm3.33$ & $2.96\pm0.98$ \\
6 & 02:25:23.34 & -04:45:53.9 & $26.55\pm0.07$ & $3.712^{+0.001}_{-0.001}$ & $-19.21\pm0.14$ & $2.54\pm0.43$ & $0.33\pm0.05$ & $7.71\pm1.66$ & $7.25\pm1.35$ \\
7 & 02:25:42.21 & -04:50:11.4 & $26.21\pm0.05$ & $3.715^{+0.001}_{-0.001}$ & $-19.62\pm0.10$ & $6.08\pm0.93$ & $0.78\pm0.12$ & $12.63\pm2.26$ & $7.51\pm1.81$ \\
8 & 02:25:52.18 & -04:51:13.7 & $25.62\pm0.03$ & $3.736^{+0.001}_{-0.001}$ & $-20.38\pm0.05$ & $21.77\pm1.18$ & $2.85\pm0.15$ & $22.74\pm1.62$ & $6.37\pm0.54$ \\
9 & 02:26:01.69 & -04:47:38.7 & $26.04\pm0.04$ & $3.742^{+0.001}_{-0.001}$ & $-19.83\pm0.08$ & $4.57\pm0.89$ & $0.60\pm0.12$ & $7.96\pm1.66$ & $6.81\pm1.61$ \\
10 & 02:25:21.96 & -04:50:39.9 & $26.93\pm0.10$ & $3.754^{+0.001}_{-0.001}$ & $-19.02\pm0.17$ & $5.14\pm0.91$ & $0.68\pm0.12$ & $19.03\pm4.62$ & $5.60\pm1.56$ \\
11 & 02:25:23.21 & -04:49:28.5 & $26.40\pm0.06$ & $3.759^{+0.001}_{-0.001}$ & $-19.52\pm0.11$ & $3.27\pm0.64$ & $0.43\pm0.08$ & $7.68\pm1.70$ & $4.70\pm5.55$ \\
12 & 02:26:02.08 & -04:52:07.2 & $26.72\pm0.08$ & $3.805^{+0.001}_{-0.001}$ & $-19.08\pm0.17$ & $8.87\pm0.96$ & $1.21\pm0.13$ & $31.99\pm6.39$ & $4.27\pm1.43$ \\
13 & 02:25:49.65 & -04:50:47.4 & $26.83\pm0.09$ & $3.818^{+0.001}_{-0.001}$ & $-19.18\pm0.16$ & $14.51\pm1.28$ & $2.00\pm0.18$ & $48.06\pm8.57$ & $7.21\pm1.33$ \\
14 & 02:25:44.86 & -04:49:51.6 & $26.75\pm0.08$ & $3.819^{+0.001}_{-0.001}$ & $-19.17\pm0.16$ & $5.55\pm0.79$ & $0.76\pm0.11$ & $18.68\pm3.96$ & $4.11\pm1.00$ \\
15 & 02:25:44.45 & -04:48:37.0 & $26.30\pm0.05$ & $3.825^{+0.001}_{-0.001}$ & $-19.59\pm0.11$ & $5.70\pm1.02$ & $0.79\pm0.14$ & $13.04\pm2.72$ & $7.71\pm1.43$ \\
16 & 02:25:40.97 & -04:49:30.9 & $26.94\pm0.10$ & $3.839^{+0.001}_{-0.001}$ & $-19.00\pm0.19$ & $9.68\pm0.80$ & $1.35\pm0.11$ & $38.41\pm7.86$ & $5.27\pm0.85$ \\
17 & 02:25:55.72 & -04:50:06.2 & $26.12\pm0.04$ & $3.854^{+0.001}_{-0.001}$ & $-19.89\pm0.09$ & $8.36\pm0.92$ & $1.18\pm0.13$ & $14.80\pm2.05$ & $4.58\pm0.72$ \\
18 & 02:25:48.95 & -04:51:29.9 & $27.07\pm0.11$ & $3.855^{+0.001}_{-0.001}$ & $-18.85\pm0.22$ & $31.52\pm1.60$ & $4.44\pm0.22$ & $145.71\pm32.86$ & $8.54\pm0.65$ \\
19 & 02:25:45.69 & -04:50:26.7 & $25.16\pm0.02$ & $3.879^{+0.001}_{-0.001}$ & $-20.68\pm0.04$ & $22.96\pm1.26$ & $3.28\pm0.18$ & $20.00\pm1.38$ & $3.44\pm0.73$ \\
20 & 02:25:33.56 & -04:49:31.9 & $27.13\pm0.11$ & $3.890^{+0.001}_{-0.001}$ & $-18.72\pm0.24$ & $3.07\pm0.71$ & $0.44\pm0.10$ & $16.26\pm5.56$ & $3.75\pm1.05$ \\
\hline
\multicolumn{10}{c}{D1GD02 (50 galaxies)} \\
21 & 02:25:42.44 & -04:51:13.2 & $26.91\pm0.09$ & $3.896^{+0.001}_{-0.001}$ & $-18.95\pm0.20$ & $6.79\pm0.66$ & $0.98\pm0.09$ & $29.23\pm6.63$ & $7.86\pm0.90$ \\
22 & 02:25:43.65 & -04:49:41.9 & $26.91\pm0.09$ & $3.897^{+0.001}_{-0.001}$ & $-19.31\pm0.15$ & $5.38\pm0.82$ & $0.78\pm0.12$ & $16.61\pm3.51$ & $3.36\pm0.75$ \\
23 & 02:25:51.29 & -04:49:26.1 & $25.98\pm0.04$ & $3.910^{+0.001}_{-0.001}$ & $-20.02\pm0.09$ & $44.70\pm2.76$ & $6.52\pm0.40$ & $72.73\pm7.50$ & $9.63\pm0.60$ \\
24 & 02:25:18.82 & -04:50:19.4 & $27.17\pm0.12$ & $3.961^{+0.001}_{-0.001}$ & $-18.78\pm0.25$ & $8.84\pm1.17$ & $1.33\pm0.18$ & $46.51\pm13.62$ & $3.92\pm0.78$ \\
25 & 02:26:00.37 & -04:51:42.6 & $26.20\pm0.05$ & $3.976^{+0.001}_{-0.001}$ & $-19.78\pm0.11$ & $17.44\pm1.68$ & $2.64\pm0.25$ & $36.61\pm5.28$ & $6.08\pm0.91$ \\
26 & 02:25:32.24 & -04:50:36.8 & $27.08\pm0.11$ & $3.979^{+0.001}_{-0.001}$ & $-18.57\pm0.30$ & $4.71\pm0.97$ & $0.72\pm0.15$ & $30.16\pm11.57$ & $4.40\pm1.84$ \\
27 & 02:26:10.02 & -04:49:55.6 & $26.40\pm0.06$ & $3.979^{+0.001}_{-0.001}$ & $-19.78\pm0.11$ & $3.75\pm0.74$ & $0.57\pm0.11$ & $7.92\pm1.77$ & $3.63\pm0.95$ \\
28 & 02:25:56.95 & -04:52:00.6 & $26.76\pm0.08$ & $4.008^{+0.001}_{-0.001}$ & $-18.88\pm0.25$ & $10.97\pm1.52$ & $1.69\pm0.23$ & $53.76\pm15.64$ & $6.23\pm1.80$ \\
29 & 02:25:30.80 & -04:50:08.0 & $26.34\pm0.06$ & $4.032^{+0.001}_{-0.001}$ & $-19.22\pm0.19$ & $14.39\pm1.34$ & $2.26\pm0.21$ & $52.69\pm11.12$ & $6.40\pm0.86$ \\
30 & 02:26:02.73 & -04:47:59.7 & $27.06\pm0.11$ & $4.038^{+0.001}_{-0.001}$ & $-19.18\pm0.20$ & $8.21\pm1.16$ & $1.29\pm0.18$ & $31.10\pm7.66$ & $9.22\pm2.40$ \\
31 & 02:25:42.96 & -04:49:06.8 & $26.65\pm0.07$ & $4.039^{+0.001}_{-0.001}$ & $-19.45\pm0.16$ & $4.91\pm0.74$ & $0.77\pm0.12$ & $14.61\pm3.19$ & $3.70\pm1.12$ \\
32 & 02:25:48.14 & -04:50:14.6 & $26.97\pm0.10$ & $4.048^{+0.001}_{-0.001}$ & $-18.98\pm0.24$ & $8.11\pm1.57$ & $1.28\pm0.25$ & $37.22\pm11.60$ & $5.40\pm2.27$ \\
33 & 02:25:51.19 & -04:49:15.3 & $26.30\pm0.05$ & $4.121^{+0.001}_{-0.001}$ & $-19.86\pm0.12$ & $16.19\pm1.20$ & $2.67\pm0.20$ & $34.34\pm4.79$ & $6.43\pm0.60$ \\
34 & 02:25:40.55 & -04:49:04.0 & $26.19\pm0.05$ & $4.131^{+0.001}_{-0.001}$ & $-19.88\pm0.12$ & $9.75\pm0.77$ & $1.62\pm0.13$ & $20.46\pm2.87$ & $3.66\pm0.49$ \\
35 & 02:25:26.36 & -04:50:34.1 & $27.00\pm0.10$ & $4.155^{+0.001}_{-0.001}$ & $-19.19\pm0.22$ & $5.97\pm0.80$ & $1.00\pm0.14$ & $23.99\pm6.36$ & $5.14\pm1.33$ \\
36 & 02:25:16.65 & -04:49:46.8 & $26.87\pm0.09$ & $4.156^{+0.001}_{-0.001}$ & $-19.23\pm0.22$ & $9.12\pm1.56$ & $1.54\pm0.26$ & $35.44\pm9.97$ & $6.45\pm1.35$ \\
37 & 02:26:09.32 & -04:51:25.2 & $26.71\pm0.08$ & $4.209^{+0.001}_{-0.001}$ & $-19.25\pm0.23$ & $5.47\pm1.03$ & $0.95\pm0.18$ & $21.43\pm6.50$ & $3.27\pm9.24$ \\
38 & 02:25:10.34 & -04:48:34.8 & $26.06\pm0.04$ & $4.250^{+0.001}_{-0.001}$ & $-20.25\pm0.10$ & $20.04\pm1.24$ & $3.56\pm0.22$ & $32.14\pm3.79$ & $11.03\pm0.78$ \\
39 & 02:25:29.32 & -04:47:49.8 & $26.40\pm0.06$ & $4.278^{+0.001}_{-0.001}$ & $-19.91\pm0.15$ & $21.27\pm1.37$ & $3.84\pm0.25$ & $47.48\pm7.55$ & $9.91\pm0.63$ \\
40 & 02:25:50.08 & -04:50:27.9 & $25.82\pm0.03$ & $4.313^{+0.001}_{-0.001}$ & $-20.14\pm0.13$ & $6.75\pm1.22$ & $1.24\pm0.22$ & $12.37\pm2.70$ & $2.55\pm1.45$ \\
41 & 02:26:06.44 & -04:49:58.8 & $25.01\pm0.02$ & $4.314^{+0.001}_{-0.001}$ & $-21.00\pm0.06$ & $15.67\pm1.92$ & $2.88\pm0.35$ & $13.01\pm1.75$ & $5.77\pm1.19$ \\
42 & 02:25:37.04 & -04:48:35.4 & $26.20\pm0.05$ & $4.318^{+0.001}_{-0.001}$ & $-19.91\pm0.15$ & $16.81\pm1.50$ & $3.10\pm0.28$ & $38.10\pm6.72$ & $5.98\pm1.09$ \\
43 & 02:25:29.80 & -04:50:37.2 & $26.25\pm0.05$ & $4.321^{+0.001}_{-0.001}$ & $-20.03\pm0.14$ & $3.68\pm0.71$ & $0.68\pm0.13$ & $7.48\pm1.76$ & $2.64\pm0.77$ \\
44 & 02:25:17.87 & -04:47:58.1 & $27.12\pm0.11$ & $4.363^{+0.001}_{-0.001}$ & $-18.93\pm0.38$ & $5.83\pm1.44$ & $1.10\pm0.27$ & $33.49\pm16.12$ & $5.23\pm1.65$ \\
45 & 02:25:21.37 & -04:46:50.2 & $26.10\pm0.04$ & $4.371^{+0.001}_{-0.001}$ & $-20.10\pm0.14$ & $24.74\pm1.26$ & $4.69\pm0.24$ & $48.50\pm7.26$ & $8.80\pm0.48$ \\
46 & 02:25:34.96 & -04:50:25.5 & $26.16\pm0.05$ & $4.436^{+0.001}_{-0.001}$ & $-19.77\pm0.21$ & $12.31\pm1.97$ & $2.42\pm0.39$ & $34.01\pm9.13$ & $6.12\pm1.65$ \\
47 & 02:25:36.93 & -04:49:27.1 & $26.09\pm0.04$ & $4.442^{+0.001}_{-0.001}$ & $-20.20\pm0.15$ & $13.10\pm1.36$ & $2.58\pm0.27$ & $24.31\pm4.29$ & $7.61\pm0.92$ \\
48 & 02:25:11.51 & -04:48:28.9 & $26.78\pm0.08$ & $4.471^{+0.001}_{-0.001}$ & $-19.12\pm0.37$ & $35.58\pm1.71$ & $7.12\pm0.34$ & $182.13\pm75.23$ & $6.68\pm0.46$ \\
49 & 02:25:11.85 & -04:50:18.0 & $27.15\pm0.12$ & $4.534^{+0.001}_{-0.001}$ & $-19.16\pm0.39$ & $13.66\pm1.83$ & $2.83\pm0.38$ & $69.58\pm31.77$ & $6.02\pm0.93$ \\
50 & 02:25:45.42 & -04:50:13.7 & $26.73\pm0.08$ & $4.638^{+0.001}_{-0.001}$ & $-19.72\pm0.30$ & $15.65\pm1.31$ & $3.41\pm0.29$ & $50.01\pm16.57$ & $5.87\pm0.77$ \\
\hline
\multicolumn{10}{c}{D4GD01 (10 galaxies)} \\
43 & 22:16:56.09 & -17:19:39.0 & $25.04\pm0.02$ & $3.649^{+0.001}_{-0.001}$ & $-20.68\pm0.04$ & $4.24\pm0.74$ & $0.52\pm0.09$ & $3.18\pm0.57$ & $4.74\pm1.11$ \\
44 & 22:16:43.36 & -17:16:37.9 & $26.13\pm0.05$ & $3.678^{+0.001}_{-0.001}$ & $-19.89\pm0.08$ & $7.18\pm1.01$ & $0.90\pm0.13$ & $11.32\pm1.81$ & $5.52\pm0.89$ \\
45 & 22:16:48.25 & -17:20:18.9 & $27.01\pm0.12$ & $3.679^{+0.001}_{-0.001}$ & $-18.72\pm0.22$ & $1.31\pm0.43$ & $0.16\pm0.05$ & $6.10\pm2.43$ & $5.10\pm11.14$ \\
46 & 22:16:48.21 & -17:21:21.8 & $26.16\pm0.05$ & $3.719^{+0.001}_{-0.001}$ & $-19.96\pm0.08$ & $9.00\pm1.15$ & $1.16\pm0.15$ & $13.67\pm2.02$ & $5.91\pm0.96$ \\
47 & 22:16:56.07 & -17:15:31.9 & $25.68\pm0.03$ & $3.736^{+0.001}_{-0.001}$ & $-20.13\pm0.07$ & $3.64\pm0.82$ & $0.48\pm0.11$ & $4.78\pm1.12$ & $3.98\pm1.88$ \\
48 & 22:17:10.34 & -17:27:41.6 & $25.55\pm0.03$ & $3.988^{+0.001}_{-0.001}$ & $-20.73\pm0.05$ & $5.20\pm1.22$ & $0.79\pm0.19$ & $4.60\pm1.10$ & $5.93\pm2.48$ \\
49 & 22:16:47.70 & -17:22:27.6 & $26.38\pm0.07$ & $4.070^{+0.001}_{-0.001}$ & $-19.58\pm0.16$ & $9.87\pm1.33$ & $1.58\pm0.21$ & $26.44\pm5.50$ & $7.73\pm1.38$ \\
50 & 22:16:47.71 & -17:21:36.2 & $27.05\pm0.12$ & $4.095^{+0.001}_{-0.001}$ & $-19.24\pm0.21$ & $5.40\pm1.12$ & $0.88\pm0.18$ & $20.04\pm6.02$ & $-2.14\pm6.22$ \\
51 & 22:16:53.40 & -17:21:26.0 & $26.30\pm0.06$ & $4.288^{+0.001}_{-0.001}$ & $-19.54\pm0.22$ & $6.38\pm1.33$ & $1.16\pm0.24$ & $20.06\pm6.08$ & $2.77\pm1.05$ \\
52 & 22:16:55.57 & -17:30:05.8 & $26.67\pm0.09$ & $4.546^{+0.001}_{-0.001}$ & $-19.76\pm0.26$ & $3.51\pm1.08$ & $0.73\pm0.22$ & $10.35\pm4.26$ & $2.70\pm1.30$ \\
\enddata
\end{deluxetable*}

\begin{figure*}
\epsscale{1.1}
\plotone{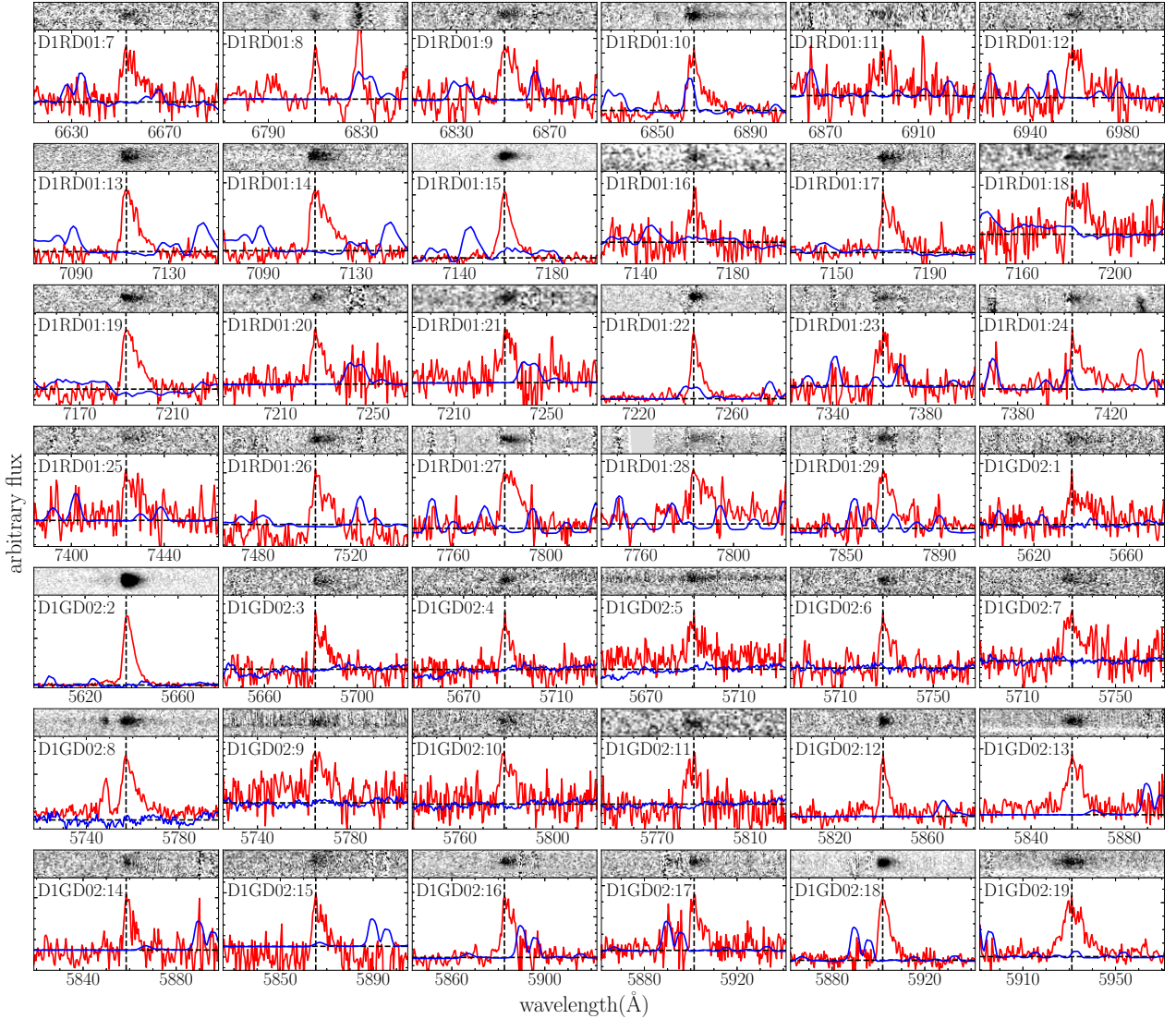}
\vspace{-3mm}
\caption{Spectra of all dropout galaxies having Ly$\alpha$ emission lines.
    The field and object IDs are indicated at the upper left corner (column 1 of Table \ref{tab:spec}).
    The vertical and horizontal dashed lines show the wavelength of Ly$\alpha$ emission and the zero level of
    flux, respectively.}
\label{fig:spec}
\end{figure*}

\setcounter{figure}{1}
\begin{figure*}
\epsscale{1.1}
\plotone{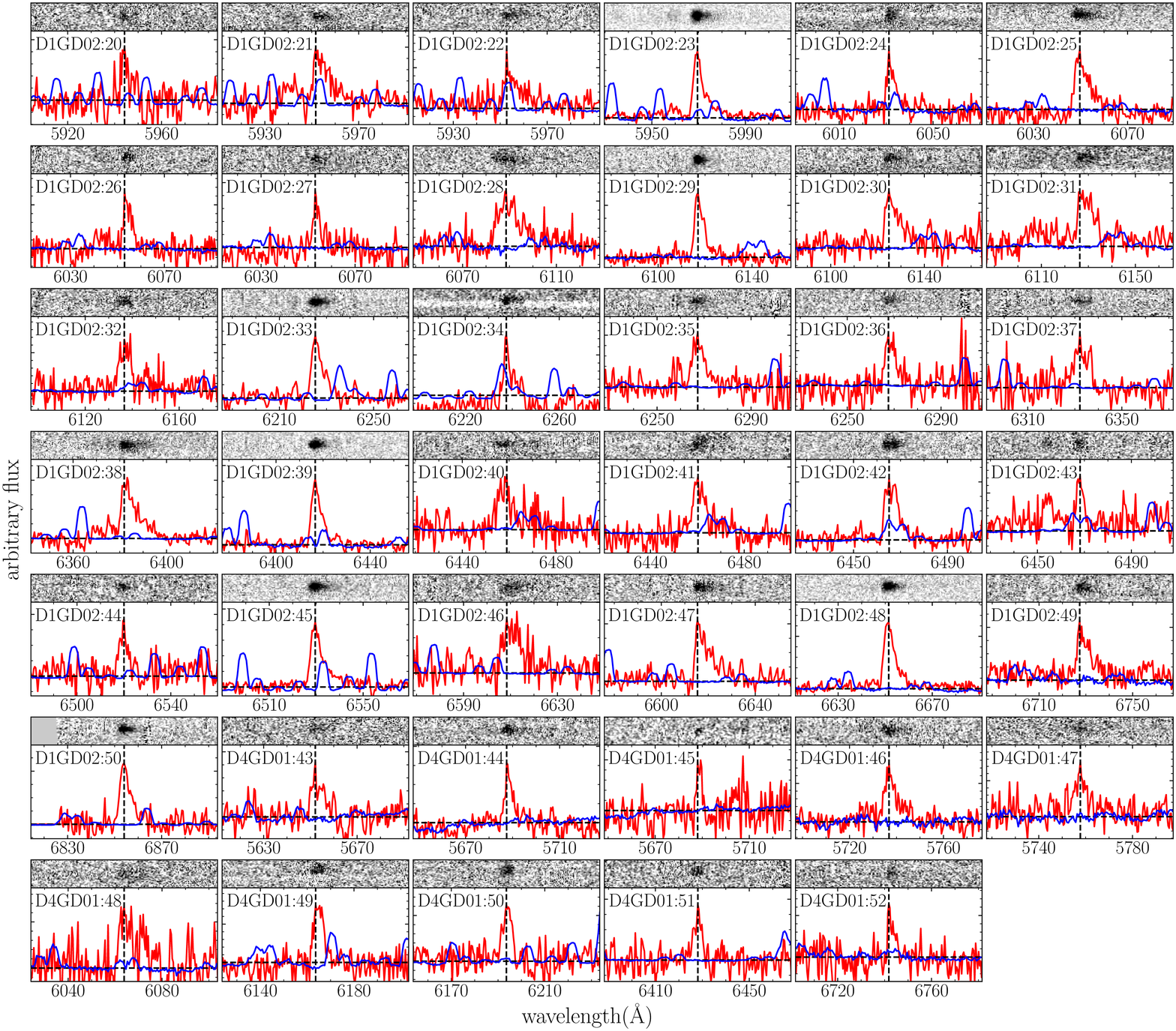}
\vspace{-3mm}
\caption{(Continued.)}
\end{figure*}

\section{Results \label{sec:res}}
In the following subsections, we will investigate whether there are protoclusters or not in each overdense region
based on the both spatial and redshift concentrations of galaxies.
At first, in Section \ref{sec:D1RD01}, \ref{sec:D1GD02}, and \ref{sec:D4GD01}, we will make a statistical test to
see whether the number of galaxies contained in a three-dimensional region is high enough to discard the
possibility that the overdensity is due to a random fluctuation. 
In this test, we set significance level at 5\%.
Then, in Section \ref{sec:sum}, we will perform a theoretical comparison in order to connect observed galaxy
concentrations to halo masses.
If the expected descendant halo masses at $z=0$ of such significant galaxy concentrations are expected to be
$>10^{14}\,\mathrm{M_\sun}$, we will be able to confirm the existence of protoclusters.
In this process of protocluster confirmation, we need to set some arbitrary criteria or assumptions.
The first is a box size to calculate three-dimensional galaxy overdensity, while a second assumption is related to
the observational bias of galaxy populations for tracing underlying structures.
We will discuss criteria and possible biases for the estimate of three-dimensional overdensity in the following.

In T16, we have estimated the expected distribution of protocluster galaxies based on the light-cone models
constructed by \citet{henriques12}.
The typical sizes of protoclusters in redshift and spatial coordinates ($L_z$, $L_\mathrm{sky}$) are found to be
$L_z\lesssim0.03\mathrm{-}0.04$ and $L_\mathrm{sky}\lesssim10\,\mathrm{arcmin}$.
The size of protoclusters is strongly dependent on the descendant mass at $z=0$; for example, the size of
progenitors of $>10^{15}\,M_\sun$ halos is about twice as large as that of $\sim10^{14}\,M_\sun$ halos.
Even for such rich protoclusters, a significant excess of galaxy density can be found with the above scale because
galaxy density in a protocluster tends to increase toward the center. 
Thus, we will estimate the strength of galaxy clustering within the three-dimensional space of $L_z\sim0.04$ and
$L_\mathrm{sky}\sim10\,\mathrm{arcmin}$ to find out protoclusters.
It should be noted that redshift can be dependent on both line-of-sight distance and radial velocity; however, the
redshift difference between protocluster members is mainly caused by their spatial separation.
Based on the light-cone model, the typical difference between apparent and geometrical redshifts is
$0.001\mathrm{-}0.004$, which is about ten times smaller than the expected protocluster size in redshift space.
Even for the progenitors of rich clusters ($>10^{15}\,\mathrm{M_\sun}$), it is $0.001\mathrm{-}0.006$.
Thus we effectively regard redshift as the parameter of radial distance.

\setcounter{figure}{3}
\begin{figure*}
\epsscale{1.0}
\plotone{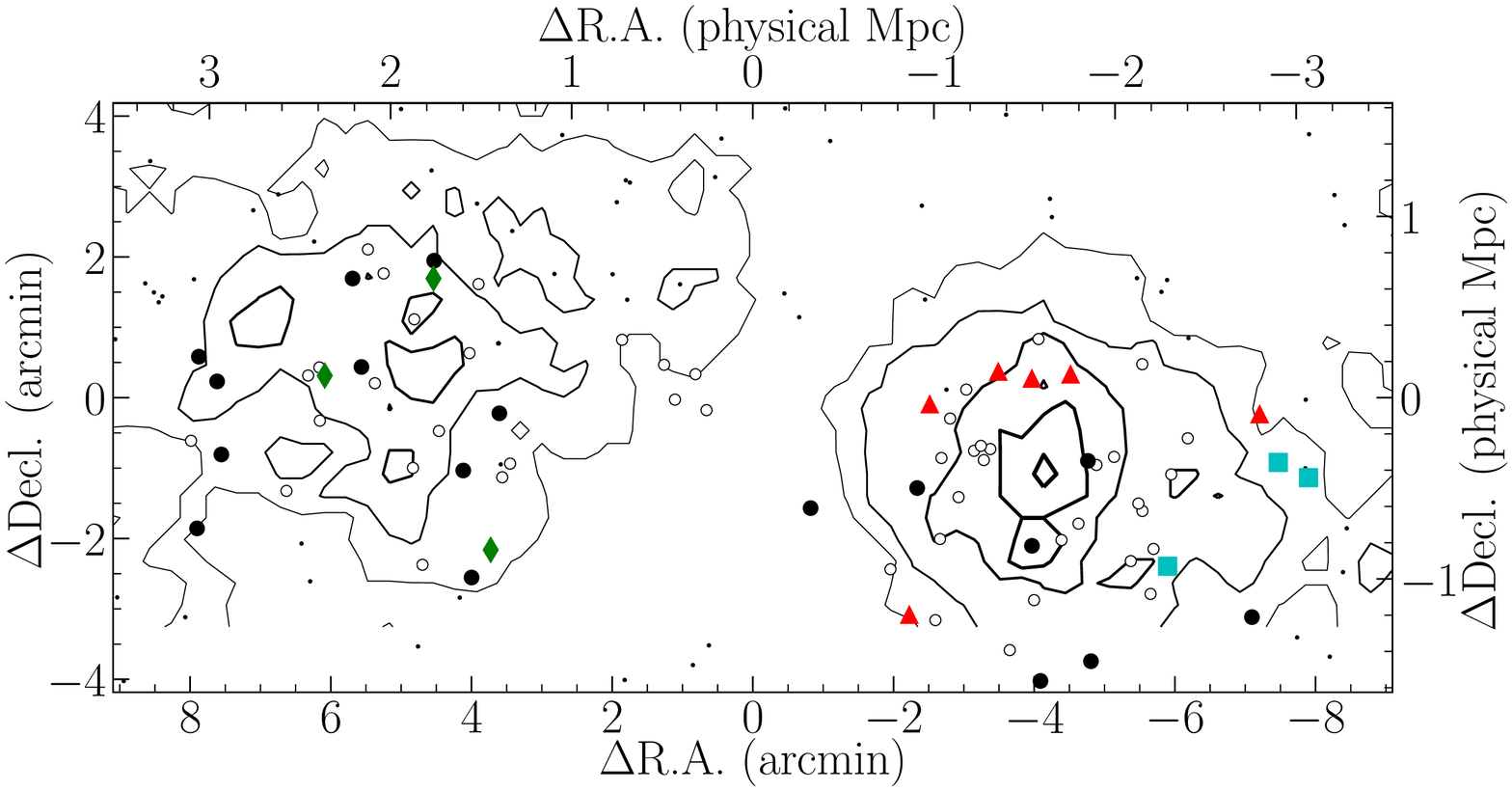}
\vspace{-3mm}
\caption{Sky distribution of $r$-dropout galaxies and number density contours in and around the D1RD01 region.
    Ly$\alpha$-detected galaxies are marked by filled symbols (red triangle: protocluster, cyan square: foreground
    group, green diamond: background group, black circle: field galaxies), and Ly$\alpha$-undetected galaxies are
    indicated by open circles.
    The dots are spectroscopically unobserved galaxies.
    The lines correspond to the contours of the surface overdensity from $4\sigma$ to $0\sigma$ in steps of
    $1\sigma$.
    The origin $(0,\,0)$ is $(\mathrm{R.A.},\,\mathrm{Decl.})=(02:25:01.89,\,-04:54:51.5)$.}
\label{fig:sky_rdrop}
\end{figure*}

\setcounter{figure}{2}
\begin{figure}
\epsscale{1.2}
\plotone{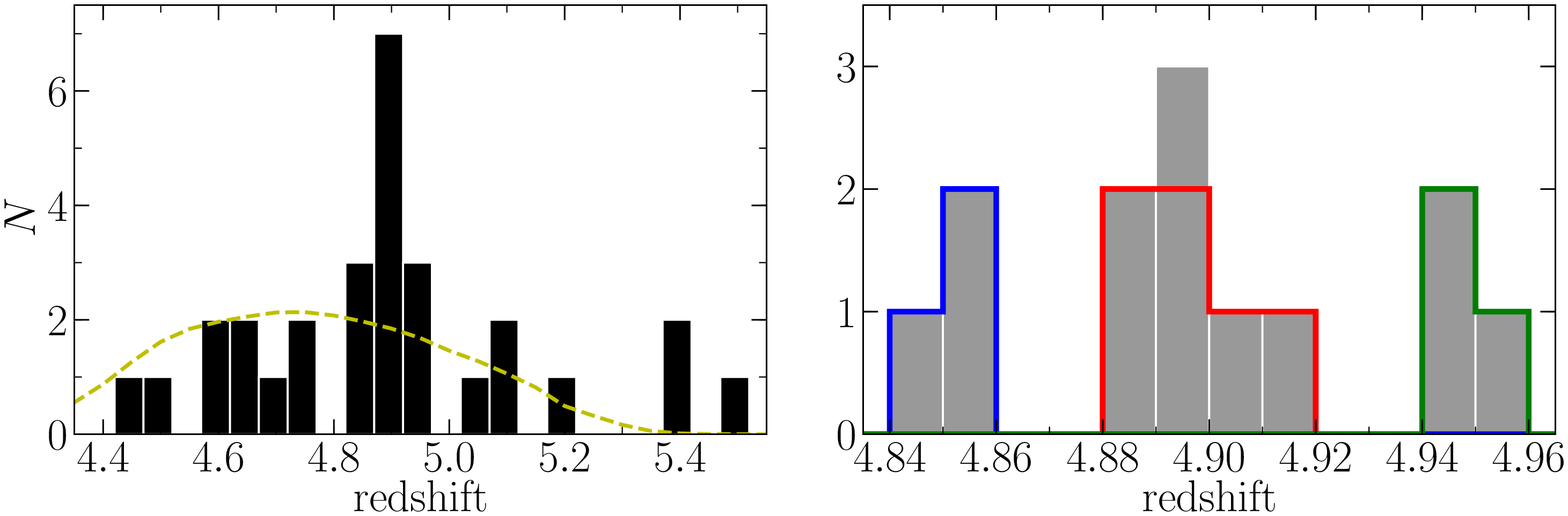}
\vspace{-7mm}
\caption{Left-hand panel: redshift distribution of 29 $r$-dropout galaxies with bin size of $\Delta z=0.05$ in and
    around the D1RD01 region.
    The yellow dashed line shows the selection function of $r$-dropout galaxies.
    Right-hand panel: close-up of the protocluster redshift range, with a bin size of $\Delta z=0.01$.
    The red, blue, and green lines indicate the galaxies of the protocluster and fore/background groups,
    respectively.}
\label{fig:z_rdrop}
\vspace{-3mm}
\end{figure}

In addition, we have to consider the possible bias of overdensity depending on galaxy population.
Since we rely on Ly$\alpha$ emission to determine the redshifts of dropout galaxies, we would miss old or dusty
galaxies, from which Ly$\alpha$ emission cannot escape easily.
Even among star-forming galaxies, their Ly$\alpha$ strength can vary widely depending on the kinematics, geometry,
and column density of interstellar medium \citep[e.g.,][]{du18,marchi19}.
As for the environmental dependence of Ly$\alpha$ emission, there are some controversial results: for example,
\citet{dey16} reported that Ly$\alpha$ luminosity is enhanced in a protocluster at $z=3.87$, while
\citet{shimakawa17} found Ly$\alpha$ depletion in a protocluster at $z=2.53$.
Furthermore, \citet{shi19b} conducted a follow-up investigation of LAE distribution around a $z=3.13$ protocluster
which were initially identified by the overdensity of dropout galaxies in T16.
They found that the peak of LAE overdensity (3.6 times higher than the average) is $\sim10\,\mathrm{arcmin}$
($\sim4.6\,\mathrm{physical\>Mpc}$) away from the protocluster, though the LAE overdense region is elongated
toward the protocluster and the protocluster itself is centered on a 1.8 times higher dense region of LAEs.
These studies suggested that the overdensity fluctuates depending on what galaxy population is used to trace it,
as expected.
However, taking into account that star-forming galaxies are dominant even in protoclusters and the strength of
Ly$\alpha$ emissions is mainly affected by galaxy internal properties rather than environments, it is feasible to
confirm the existence of protoclusters by the combination of overdensity of dropout galaxies and follow-up
spectroscopy of Ly$\alpha$ emissions.
It should be noted that such protoclusters would be only a subsample of all protoclusters, and confirmed
protocluster galaxies themselves are only a subset of all the members in a protocluster.
Deep multi-wavelength observation and complete spectroscopy are necessary to reveal a complete sample of
protoclusters, and this is beyond the scope of this study.
The results of follow-up observations on each overdense region are described in the following subsections, and
Table \ref{tab:pcl} summarizes the results of our protocluster confirmation.

\subsection{The Protocluster at $z\sim4.9$ in the D1 Field \label{sec:D1RD01}}
In the previous work of T16, it was not clear whether a protocluster exits in the overdense region of D1RD01 or
not because the total number of redshift identifications was only six, which was too small to draw a firm
conclusion.
However, two of them are tightly clustered at $z=4.89$.
In this study, we have increased the number of confirmed galaxies by a factor of five.
As shown in the updated redshift distribution of Figure \ref{fig:z_rdrop}, there is a significant peak at
$z\sim4.9$.
As the FoV of DEIMOS ($16.3\times5.0\,\mathrm{arcmin^2}$) is larger than the typical size of protoclusters
($\sim5\,\mathrm{arcmin}$ radius; see Figure 8 of T16), we focus on the part of the FoV of DEIMOS including the
peak of overdensity.
Six galaxies, with ID=5, 6, 15, 17, 18, and 19, are tightly clustered in both redshift and spatial coordinates
($\Delta z=0.025$ at $z=4.898$ and $3.5\,\mathrm{arcmin}$ radius from the center of
(R.A., Decl.)=(02:24:47.03, -04:54:43.3)).
Although the redshift of ID16 is within the redshift range of these six galaxies, its sky position is
$>10\,\mathrm{arcmin}$ away from these six galaxies.
We have estimated the probability of finding this clustering structure by chance if galaxies are randomly
distributed according to the redshift selection function of $r$-dropout galaxies with the following method.
In the overdense region, where six clustering galaxies are located (the area of $\Delta\mathrm{R.A.}<0$ in Figure
\ref{fig:sky_rdrop}), there are 16 galaxies including fore/background galaxies.
Using the selection function of $r$-dropout galaxies, the same number of galaxies as observed (16 galaxies) are
randomly distributed in redshift.
Then, we check whether more than six galaxies are clustering within $\Delta z\leq0.025$.
We have repeated this random realization 10,000 times, and the probability is found to be $<0.2\%$ ($>3.1\sigma$
significance).
Since the clustering structure cannot be attributed to just a random coincidence, these six galaxies are highly
expected to be physically related.
This is the evidence for the existence of a protocluster in the overdense region of D1RD01.
The close galaxy pair found by T16 turns out to be part of this protocluster.
These six galaxies are indicated in red in Figures \ref{fig:z_rdrop} and \ref{fig:sky_rdrop}, respectively.

\setcounter{figure}{5}
\begin{figure*}
\epsscale{1.0}
\plotone{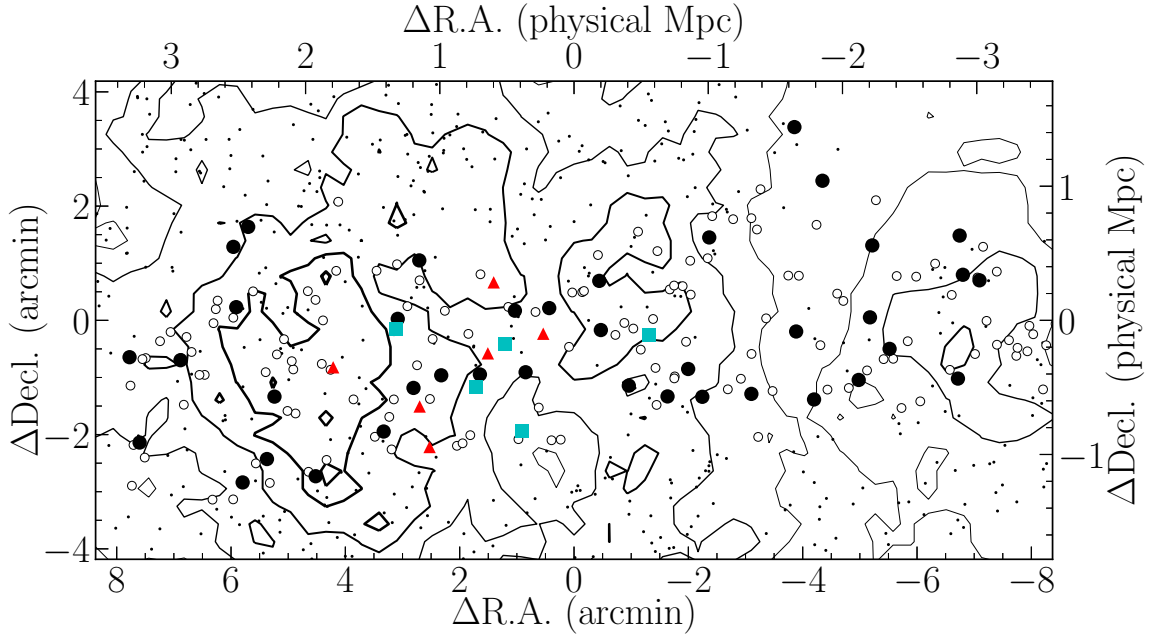}
\vspace{-3mm}
\caption{Sky distribution of $g$-dropout galaxies and number density contours in and around the D1GD02 region.
    Ly$\alpha$-detected galaxies are marked by filled symbols (red triangle: protocluster, cyan square:
    background groups, black circle: field galaxies), and Ly$\alpha$-undetected galaxies are indicated by open
    circles.
    The dots are spectroscopically unobserved galaxies.
    The lines correspond to the contours of the surface overdensity from $4\sigma$ to $0\sigma$ in steps of
    $1\sigma$.
    The origin $(0,\,0)$ is $(\mathrm{R.A.},\,\mathrm{Decl.})=(02:25:38.81,\,-04:49:16.9)$.}
\label{fig:sky_gdrop1}
\vspace{-1mm}
\end{figure*}

\setcounter{figure}{4}
\begin{figure}
\epsscale{1.2}
\plotone{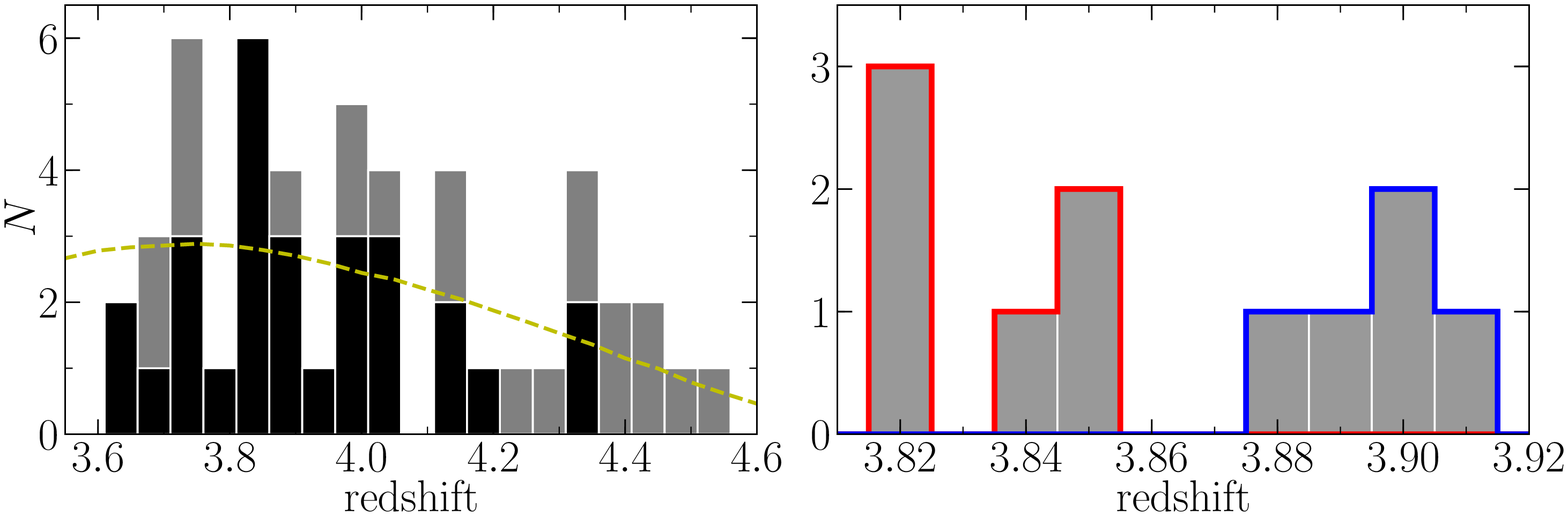}
\vspace{-7mm}
\caption{Left-hand panel: redshift distribution of 50 $g$-dropout galaxies with bin size of $\Delta z=0.05$ in and
    around the D1GD02 region.
    The gray histogram shows all 50 galaxies over the whole FoV of DEIMOS, and the black one shows only those
    within the typical size of protoclusters from the overdensity peak (the area of
    $\Delta\mathrm{R.A.}>0\,\mathrm{arcmin}$ in Figure \ref{fig:sky_gdrop1}).
    In the black histogram, we find a significant peak at $z\sim3.8$, and peaks at $z\sim3.7$ or 4.0 shown by the
    gray histogram turn out to be incidental peaks due to the wider FoV than protocluster size.
    The yellow dashed line shows the selection function of $g$-dropout galaxies.
    Right-hand panel: close-up of the protocluster redshift range, with a bin size of $\Delta z=0.01$.
    The red and blue lines indicate the galaxies of the protocluster and background group, respectively.}
\label{fig:z_gdrop1}
\vspace{-3mm}
\end{figure}

We have also found two galaxy groups which are located closely around the protocluster (indicated by blue and
green in Figures \ref{fig:z_rdrop} and \ref{fig:sky_rdrop}).
Both groups consist of three galaxies, and their redshift and spatial sizes are less than $\Delta z=0.015$ and
$2.4\,\mathrm{arcmin}$ radius.
Although such a grouping containing three galaxies within $\Delta z=0.015$ can be reproduced by random
distributions with a probability of $20\%\mathrm{-}37\%$ ($0.9\sigma\mathrm{-}1.3\sigma$), we have found two
nearby groups simultaneously.
Furthermore, they are surprisingly arranged in the foreground and background of the protocluster, as if they form
a filamentary structure in the large-scale structure of the universe.
The redshift separation between the protocluster and the fore/background groups is only 0.05, corresponding to
$4.6\,\mathrm{Mpc}$ in physical scale.
These foreground (blue) and background (green) groups are composed of ID=4, 13, and 14 and ID=20, 21, and 22,
respectively.
The whole redshift range including the protocluster and fore/background groups is $\Delta z=0.12$.
We have formally applied the above calculation of the significance of clustering to this large system, and it is
found to be 3.3\% ($2.1\sigma$).
It should be noted that, if there is no neighboring groups, the galaxy distribution of six galaxies within
$\Delta z=0.12$ can be reproduced from random distribution with the probability of 46\%.
Thus, the fore/background groups are essential components, suggesting the existence of a large-scale structure
composed of the protocluster and fore/background groups though the statistical significance is $2\sigma$ level.
As the FoV of our follow-up spectroscopy is limited to a part of the surrounding area of the protocluster, we need
more follow-up observations to cover other surrounding areas, which may result in the finding of other neighboring
groups.
It should be noted that, due to the wide redshift window of dropout selection, the spatial clustering of dropout
galaxies embedded in a projected overdense region might favor elongated large-scale structure which points toward
us.

\subsection{The Possible Protocluster at $z\sim3.8$ in the D1 Field \label{sec:D1GD02}}
The overdense region of D1GD02 is newly observed in this study, and 50 galaxies are spectroscopically confirmed.
From the redshift distribution of all galaxies as shown in Figure \ref{fig:z_gdrop1}, it is difficult to find a
clear redshift spike.
However, when we focus only on the typical size of protoclusters including the peak of overdensity (the area of
$\Delta\mathrm{R.A.}>0\,\mathrm{arcmin}$ in Figure \ref{fig:sky_gdrop1}), there is a peak at $z\sim3.8$ indicated
by the black histogram in the left panel of Figure \ref{fig:z_gdrop1}.
The redshift spike consists of six galaxies (ID=13-18), ranging over $\Delta z=0.036$ centered at $z=3.834$.
As shown by red points in Figure \ref{fig:sky_gdrop1}, these six galaxies are also closely clustered in spatial
coordinates ($\sim2\,\mathrm{arcmin}$ radius).
In the same way as we did for $r$-dropout galaxies, we find that such a clustering structure can be reproduced by
random distribution with a probability of 5.9\% ($1.9\sigma$).
Based on this probability, the clustering signature of these six galaxies is likely to result from the existence
of a protocluster.
If these six galaxies form a single protocluster, it should be noted that this protocluster seems to exhibit a
bimodal redshift distribution (the red histogram in the right panel of Figure \ref{fig:z_gdrop1}).
In addition, there is another group including five galaxies (ID=19-23) around $z=3.895$ just behind the
protocluster at $z=3.834$ (the blue histogram in the right panel of Figure \ref{fig:z_gdrop1}).
The redshift width of these five galaxies is $\Delta z=0.032$, and the probability that five galaxies happen to be
located within this redshift width by random distribution is $\sim20\%$ ($\sim1.3\sigma$).
This probability is not small enough to deny that the background group might be just an apparent clustering, not
physically associated with each other.
However, it would be unlikely that these two galaxy groups are closely located by chance because the expected
redshift distribution of dropout galaxies is $\Delta z\sim1$, which is much larger than their redshift separation
of $\Delta z=0.061$.
The probability of the reproduction of this large system including eleven galaxies within $\Delta z=0.10$ is
estimated to be 8.5\% ($1.7\sigma$) by a random distribution. 
In the overdense region of D1GD02, we have found a possible protocluster at $z=3.834$ with a moderate ($1.9\sigma$)
significance level, which does not allow us to make a definitive identification of the overdense region as a
protocluster.
We will need more spectroscopic identifications in order to definitely confirm this as a protocluster and reveal
the large-scale structure around it. 

\setcounter{figure}{6}
\begin{figure}
\epsscale{1.2}
\plotone{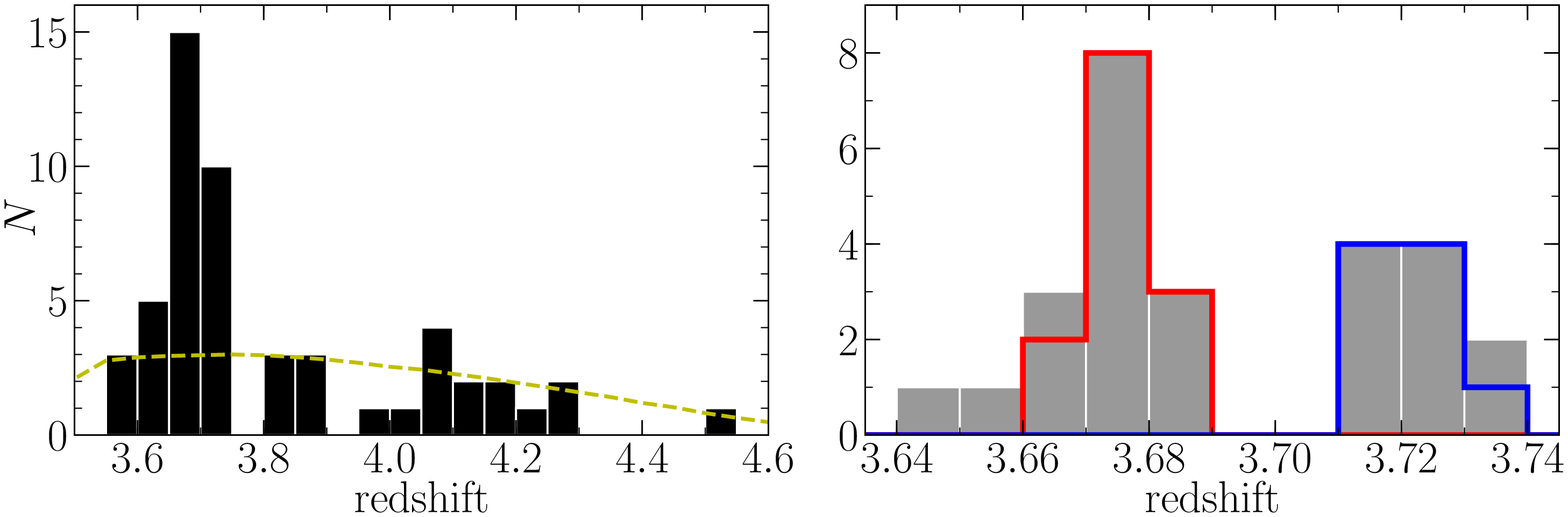}
\vspace{-7mm}
\caption{Left-hand panel: redshift distribution of 52 $g$-dropout galaxies and an AGN with bin size of
    $\Delta z=0.05$, in and around the D4GD01 region.
    The yellow dashed line shows the selection function of $g$-dropout galaxies.
    Right-hand panel: close-up of the protocluster redshift range with a bin size of $\Delta z=0.01$.
    The red and blue lines indicate the galaxies of the main and background protoclusters, respectively.}
\label{fig:z_gdrop2}
\vspace{-3mm}
\end{figure}

\begin{figure}
\epsscale{1.2}
\plotone{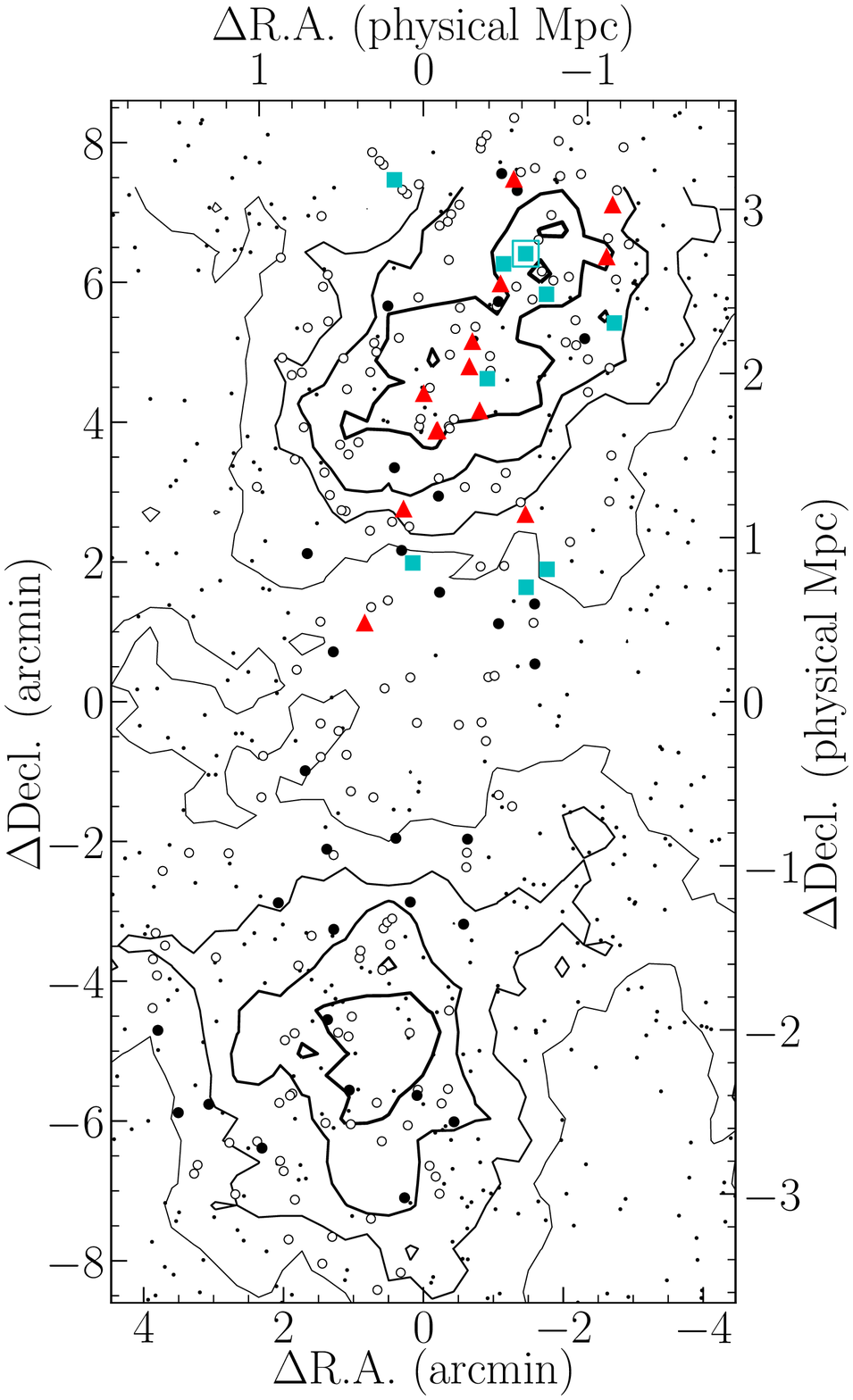}
\vspace{-8mm}
\caption{Sky distribution of $g$-dropout galaxies and number density contours in and around the D4GD01 region.
    Ly$\alpha$-detected galaxies are marked by filled symbols (red triangle: main protocluster, cyan square:
    background protocluster, black circle: field galaxies), and Ly$\alpha$-undetected galaxies are indicated by
    open circles.
    The dots are spectroscopically unobserved galaxies, and the cyan double square is the AGN.
    The lines correspond to the contours of the surface overdensity from $4\sigma$ to $0\sigma$ in a step of
    $1\sigma$.
    The origin $(0,\,0)$ is $(\mathrm{R.A.},\,\mathrm{Decl.})=(22:16:54.38,\,-17:22:59.9)$.}
\label{fig:sky_gdrop2}
\vspace{-3mm}
\end{figure}

\begin{deluxetable*}{cccccc}
\tabletypesize{\normalsize}
\tablecaption{Results of the protocluster confirmation \label{tab:pcl}}
\tablewidth{0pt}
\tablehead{Name & R.A.\tablenotemark{a} & Decl.\tablenotemark{a} & redshift\tablenotemark{a} &
            $N_\mathrm{mem}$\tablenotemark{b} & $\sigma_v$ ($\mathrm{km\,s^{-1}}$)}
\startdata
D1RD01 & 02:24:47.03 & -04:54:43.3 & 4.898 & 6 & $502.6\pm171.2$ \\
D1GD02 & 02:25:46.90 & -04:50:02.5 & 3.834 & 6 & $1025.0\pm393.5$ \\
D4GD01 & 22:16:51.37 & -17:18:24.6 & 3.675 & 13 & $329.2\pm73.3$ \\
D4GD01-back & 22:16:48.16 & -17:17:47.0 & 3.721 & 9 & $229.1\pm129.9$ \\
\enddata
\tablenotetext{a}{Biweight location of protocluster members.}
\tablenotetext{b}{Number of protocluster members.}
\vspace{-9mm}
\end{deluxetable*}

\subsection{The Protoclusters at $z\sim3.7$ in the D4 Field \label{sec:D4GD01}}
A protocluster in the overdense region of D4GD01 was originally discovered at $z=3.67$ by T16.
This study increases the number of spectroscopically-confirmed member galaxies for a more detailed investigation.
Out of ten newly confirmed galaxies, two are found to be in the protocluster; thus, there are at least 13 member
galaxies (ID=10-20, 44, and 45), which are indicated by red in Figures \ref{fig:z_gdrop2} and \ref{fig:sky_gdrop2}.
These 13 member galaxies are in the redshift range of $\Delta z=0.016$ centered at $z=3.675$.
In none of our 10,000 randomly simulated realizations, such a clustering structure was reproduced.
In addition, we can notice that there is another clustering structure at $z=3.721$, i.e. the background of the
protocluster, which is composed of nine galaxies within $\Delta z=0.020$ in total: eight galaxies (ID=21-26, 46,
and 47), and an AGN.
This background structure can also not be explained by a random distribution, and has a comparable number of
member galaxies to other known high-redshift protoclusters \citep[e.g.,][]{ouchi05,cucciati14,lemaux18}.
Therefore, by the further follow-up spectroscopy in this study, we have not only increased the number of member
galaxies in the known protocluster at $z=3.675$, but also discovered another protocluster just behind it.
As shown in Figures \ref{fig:z_gdrop2} and \ref{fig:sky_gdrop2}, these two protoclusters are near each other
($\Delta z=0.046$, corresponding to $7.4\,\mathrm{Mpc}$ in physical scale).
These two protoclusters are expected to form a large system because we cannot reproduce such a galaxy distribution
from random realizations.

\subsection{Summary of Protocluster Confirmation \label{sec:sum}}
Based on these follow-up spectroscopy, we newly confirm two protoclusters at $z=4.898$ and 3.721, and two member
galaxies are additionally found in the known protocluster at $z=3.675$,
Furthermore, the overdense region of D1GD02 may also include a protocluster at $z=3.834$ though its statistical
significance is marginal.
We have estimated the three-dimensional galaxy overdensity for these four protoclusters, including a possible one,
by comparing with the other fore/background galaxies as field counterparts.
As both protocluster and field galaxies are selected from the same photometric sample and spectroscopically
observed in the same masks, there should be little observational bias.
However, it should be noted that protocluster galaxies could have different physical properties from field
galaxies, which could causes different LAE fractions among dropout galaxies between protoclusters and field.
In this study, we assume the same LAE fractions in the protoclusters and field.
Thus, the three-dimensional galaxy overdensity, $\delta_\mathrm{gal}$ ($=n/\bar{n}$, where $n$ and $\bar{n}$ are
the number density in a protocluster and field respectively), of protoclusters at $z=4.898$, 3.834, 3.721, and
3.675 are found to be $\delta_\mathrm{gal}=6.0^{+3.6}_{-2.4}$, $3.7^{+2.2}_{-1.5}$, $4.5^{+2.0}_{-1.5}$, and
$6.4^{+2.4}_{-1.7}$, respectively.

Three-dimensional density enables us to estimate descendant halo mass at $z=0$ by using theoretical models.
As shown in Section \ref{sec:obs}, 76\% of $>4\sigma$ overdense regions are expected to grow into galaxy clusters
with the halo mass of $>10^{14}\,\mathrm{M_\sun}$ comparing with a theoretical model \citep{henriques12}.
In this model comparison, there are 84 $>4\sigma$ overdense regions, of which 82 regions show three-dimensional
galaxy concentrations with $\delta_\mathrm{gal}>2$.
It should be noted that such galaxy concentrations can be buried in lower surface dense regions.
Since our objective of this model comparison is to predict the descendant halo mass of the observed protoclusters,
we have focused on three-dimensional galaxy concentrations embedded in $>4\sigma$ overdense regions in the
same manner as our observations.
Then, we can find the relation between three-dimensional overdensity and descendant halo mass.
The result is that protoclusters with $\delta_\mathrm{gal}=3.7$ and $4.5$ are expected to be
$(1.0\mathrm{-}5.0)\times10^{14}\,\mathrm{M_\sun}$ and $(2.2\mathrm{-}5.1)\times10^{14}\,\mathrm{M_\sun}$ halos
(the range between upper and lower quartiles).
As for protoclusters with $\delta_\mathrm{gal}>6$, only two comparable regions are identified in the theoretical
model, and their descendant halo masses are $4.0\times10^{14}\,\mathrm{M_\sun}$ or
$1.2\times10^{15}\,\mathrm{M_\sun}$.
While descendant halo mass at $z=0$ generally tends to be proportional to galaxy density at high redshifts, there
is still a large dispersion.
From this theoretical comparison, the three-dimensional galaxy concentrations identified by this study are found
to have large overdensity enough to grow into galaxy clusters ($>10^{14}\,\mathrm{M_\sun}$ halos) by $z=0$.
Therefore, we have concluded that the three overdense regions at $z=4.898$, 3.721, and 3.675 are genuine
protoclusters, while the overdense region at $z=3.834$ is still a possible candidate of a protocluster due to the
small number of confirmed galaxies.

\section{Discussion \label{sec:disc}}
\subsection{Formation of Superclusters}
As shown in Section \ref{sec:res}, the two protoclusters at $z=4.898$ and 3.675 are accompanied by neighboring
groups/protocluster.
In the context of the hierarchical structure formation model, it is expected that galaxy clusters are formed from
clumps of galaxy groups or smaller structures through repeated halo mergers.
Thus, at high redshift, we would observe some groups around a main progenitor.
Furthermore, galaxy clusters themselves frequently reside in larger-scale high-density regions, which include some
clusters, groups, or filamentary distributions of galaxies.
We can see much larger structures beyond the scale of galaxy clusters in the local universe, which are called
superclusters \citep[e.g.,][]{bahcall84}.
It is a question how the protoclusters we found at $z=4.9$ and 3.7 will evolve by $z=0$: will they become a single
rich galaxy clusters by incorporating their neighboring groups/protocluster, or will each of them develop into an
independent halo as a part of a supercluster?
The separation between the main protoclusters and their surrounding groups/protocluster is $\Delta z\sim 0.05$,
corresponding to $\sim5\,\mathrm{Mpc}$ and $\sim8\,\mathrm{Mpc}$ in physical scale at $z=4.9$ and 3.7 respectively.
The size of protoclusters depends significantly on the descendant halo mass at $z=0$.
\citet{chiang13} estimated an effective radius of protoclusters, in which 40\% of the total mass of a protocluster
is distributed, based on $N$-body dark matter simulations \citep{springel05}.
A typical size of the effective radius is $\sim1\,\mathrm{Mpc}$ in physical scale at $z\sim4\mathrm{-}5$ for the
progenitors of $1\mathrm{-}3\times10^{14}\,\mathrm{M_\sun}$ halos.
Even for those of $>10^{15}\,\mathrm{M_\sun}$ halos, the size is $\lesssim2\,\mathrm{Mpc}$.
Similarly, \citet{muldrew15} also investigated the size of protoclusters at high redshifts based on the stellar
mass of protocluster members.
They predicted that 90\% of the stellar mass of all protocluster members is on average enclosed in
$\sim2\,(4)\,\mathrm{Mpc}$ in physical scale at $z\sim4\mathrm{-}5$ for the descendants of
$1\mathrm{-}6\times10^{14}\,(>10^{15})\,\mathrm{M_\sun}$ halos by the combination of $N$-body dark matter
simulation and a semi-analytic galaxy formation model \citep{guo11}.
Based on these theoretical predictions, only if a descendant halo mass at $z=0$ is $>10^{15}\,\mathrm{M_\sun}$,
the main protoclusters and their neighboring groups/protocluster have the potential to merge into a single cluster
by $z=0$.
However, according to \citet{toshikawa18}, which used the same method to search for protoclusters as this study,
the typical descendant halo mass of $z\sim4$ protoclusters is expected to be
$\sim4\mathrm{-}8\times10^{14}\,M_\sun$ at $z=0$ based on clustering analysis and abundance matching.
Assuming that the protoclusters in this study have similar descendant halo mass, the separation between the
main protoclusters and their neighboring groups/protocluster is larger than those theoretical expectations of
typical protocluster size.
This suggests that the neighboring groups/protocluster grow into independent halos from the main protocluster.

However, we should consider the possibility that there is no physical relation between the protoclusters and the
neighboring groups (at least not at the supercluster scale), because the redshift separation between the main
protoclusters and neighboring groups/protocluster is only $\Delta z\sim0.05$.
This is much smaller than the redshift window of dropout selection ($\Delta z\sim1$).
Especially, as for the $z=4.898$ protocluster, two galaxy groups are simultaneously found at foreground and
background.
Thus, the proximity of galaxy groups would result from the underlining large-scale structure of the universe.
As about half of clusters are in superclusters at $z\lesssim0.5$ \citep[e.g.,][]{bahcall84,chow14}, some parts of
protoclusters are expected to reside in a primordial superstructure.
In the local universe ($z\lesssim0.5$), superclusters are typically $\sim20\,\mathrm{physical\>Mpc}$ in size
between the maximally separated pair of clusters in a supercluster, and the largest ones have nearly
$100\,\mathrm{physical\>Mpc}$ length, based on the extended ROSAT-ESO Flux-Limited X-ray Galaxy Cluster survey
data \citep{chon13}.
Most of such superclusters are composed of two or three galaxy clusters, and a few of them include nearly ten
clusters.
\citet{alpaslan14} also investigated the large-scale structure of the universe by using the Galaxy And Mass
Assembly survey, and a typical length of filamentary large-scale structures is found to be $\sim20\,\mathrm{Mpc}$
including eight galaxy groups.
Although the spatial size of superclusters depends on the definition or method to search, it is typically a few
tens Mpc in the local universe.
If our large systems are already detached from Hubble flow, their expected spatial sizes at $z=0$ are less than
$10\,\mathrm{Mpc}$, which is smaller than that of typical local superclusters.
In case the separations between the protoclusters and accompanying groups/protocluster are increasing according to
the Hubble flow, their expected separations at $z=0$ are $\sim30\,\mathrm{Mpc}$, comparable to local superclusters.
Therefore, the large systems including the protoclusters and accompanying groups/protocluster would be the
progenitors of superclusters rather than the chance alignment of totally unrelated groups/protoclusters.
At $z=4.9$ and 3.7, we have revealed primordial superclusters with comprehensive follow-up spectroscopy by
Keck/DEIMOS, which has a larger FoV than the typical protocluster size.
In particular, in the D4GD01 overdensity, we have found a close pair of protoclusters, whose separation is
$7.5\,\mathrm{Mpc}$ in physical scale.
According to the two-point correlation function of protocluster candidates at $z\sim4$ \citep{toshikawa18}, the
expected number of protoclusters within $\sim8\,\mathrm{Mpc}$ in physical scale from another protocluster can be
estimated to be $\sim0.20\mathrm{-}0.45$.
Therefore, if the protoclusters identified in this study are comparable to typical ones (the progenitors of
$\sim3\times10^{14}\,\mathrm{M_\sun}$ halos, not $>10^{15}\,\mathrm{M_\sun}$ halos at $z=0$), the large systems
including the protoclusters and neighboring groups would trace primordial large-scale structures instead of
multiple progenitors on a same halo merger tree.
To predict the fate of these large structures, we need to map out three-dimensional galaxy distribution more
precisely by more spectroscopic follow-up observations.

There are other examples of such large-scale structures at $z\sim2\mathrm{-}6$
\citep[e.g.,][]{ouchi05,kuiper12,dey16,topping16,cucciati18}.
Especially, \citet{cucciati18} discovered a clear example of a primordial supercluster at $z=2.45$ by
identifying seven galaxy groups within a volume of $\sim60\times60\times150\,\mathrm{comoving\,Mpc^3}$.
On the other hand, the large-scale structure in the D1RD01 region includes the protocluster and two groups over a
volume of $\sim33\times12\times64\,\mathrm{comoving\,Mpc^3}$, and, in the D4GD01 regions, the two protoclusters
are embedded in a volume of $\sim15\times10\times50\,\mathrm{comoving\,Mpc^3}$.
When the size of the primordial supercluster of \citet{cucciati18} is limited to that of the ones we found, only
two or three among the seven galaxy groups identified as the components of the primordial supercluster of
\citet{cucciati18} can be reconfirmed as its components.
Since this number is comparable to ours, we would observe a portion of primordial superclusters like one found by
\citet{cucciati18}.
\citet{topping18} also closely investigated a large-scale structure around the SSA22 protocluster at $z=3.09$.
They found two galaxy concentrations within a volume of $\sim12\times14\times43\,\mathrm{comoving\,Mpc^3}$, which
are predicted to be the size of two separated halos at $z=0$.
Our findings of the large-scale structures at $z=4.898$ and 3.675 seem to exhibit a consistent size and number of
components with the clear examples at lower redshifts.
However, according to the theoretical comparison in \citet{topping18}, the occurrence rate of such a large-scale
structure around the SSA22 protocluster is expected to be $7.4\,\mathrm{Gpc^{-3}}$.
On the other hand, the total survey volume of our $g$- and $r$-dropout galaxies over the CFHTLS Deep Fields
($\sim4\,\mathrm{deg^2}$) is only $\sim0.06\,\mathrm{Gpc^3}$.
Since \citet{topping18} predicted that two galaxy concentrations grow into the cluster pair of
$>10^{15}\,\mathrm{M_\sun}$ and $>10^{14}\,\mathrm{M_\sun}$ halos at $z=0$, our findings of the large-scale
structures may be composed of the progenitors of smaller clusters.

Although we cannot statistically calculate what fraction of protoclusters are in primordial superclusters or
isolated due to the small and heterogeneous sample of protoclusters, these findings, at least, indicate that some
components of a supercluster are simultaneously formed at high redshifts despite being a few Mpc away from each
other.
A similar result is found in a supercluster at $z\sim0.9$, in which three clusters and five groups
($4.6\mathrm{-}0.3\times10^{14}\,\mathrm{M_\sun}$ halos) are contained over $\sim20\,\mathrm{comoving\>Mpc}$ area
\citep{lemaux12}.
Considering the limited volume of our observation of the primordial superclusters, this supercluster at $z=0.9$
would be comparable to our findings; at least, it is worth to compare with a well-known supercluster at $z\sim1$.
This shows that superclusters can be formed around not only rich clusters but also typical ones.
Furthermore, \citet{hayashi19} newly found some components of this superclusters which are
$\sim50\,\mathrm{comoving\,Mpc}$ away from each other at maximum, and investigated stellar ages of red-sequence
galaxies in each component based on composite spectra.
Although they are $\sim10\mathrm{-}50\,\mathrm{comoving\,Mpc}$ away from each other, they are found to have
similar mean stellar ages, which possibly indicate that each component is formed at almost the same redshift.
Thus, we would be able to identify multiple protoclusters/groups in the progenitors of superclusters at even
higher redshifts as shown in this study or \citet{cucciati18}.
This is qualitatively predicted by the hierarchical structure formation model; however, each path of structure
formation would have a large variation due to repeated halo mergers.
Therefore, the direct observation of developmental stages will provide us with an important constraint on such 
stochastic processes.
The $z=4.898$ protocluster brings two small groups while the primordial superclusters at $z=3.675$ in the D4GD01
overdense region are composed of two protoclusters, which include almost comparable numbers of member galaxies.
The number of components in a primordial supercluster and the ratio of galaxy number between each component would
be hints to understand the formation of superclusters or the large-scale structure of the universe.
It should be noted that our follow-up spectroscopy focuses on the overdensity peak.
Surrounding regions are sparsely observed, and the spatial distribution of galaxies is traced only by dropout
galaxies having Ly$\alpha$ emissions. 
This incompleteness of observed surrounding area and galaxy population may cause the apparent difference between
at $z=4.898$ and 3.675 due to a bias against large-scale galaxy distribution or an oversight of small accompanying
groups.
Although we need further follow-up observations to cover the whole structure and to investigate galaxy population
in protoclusters, we have directly observed (a part of) primordial superclusters at $z=4.9$ and 3.7.

\begin{figure*}
\epsscale{1.1}
\plottwo{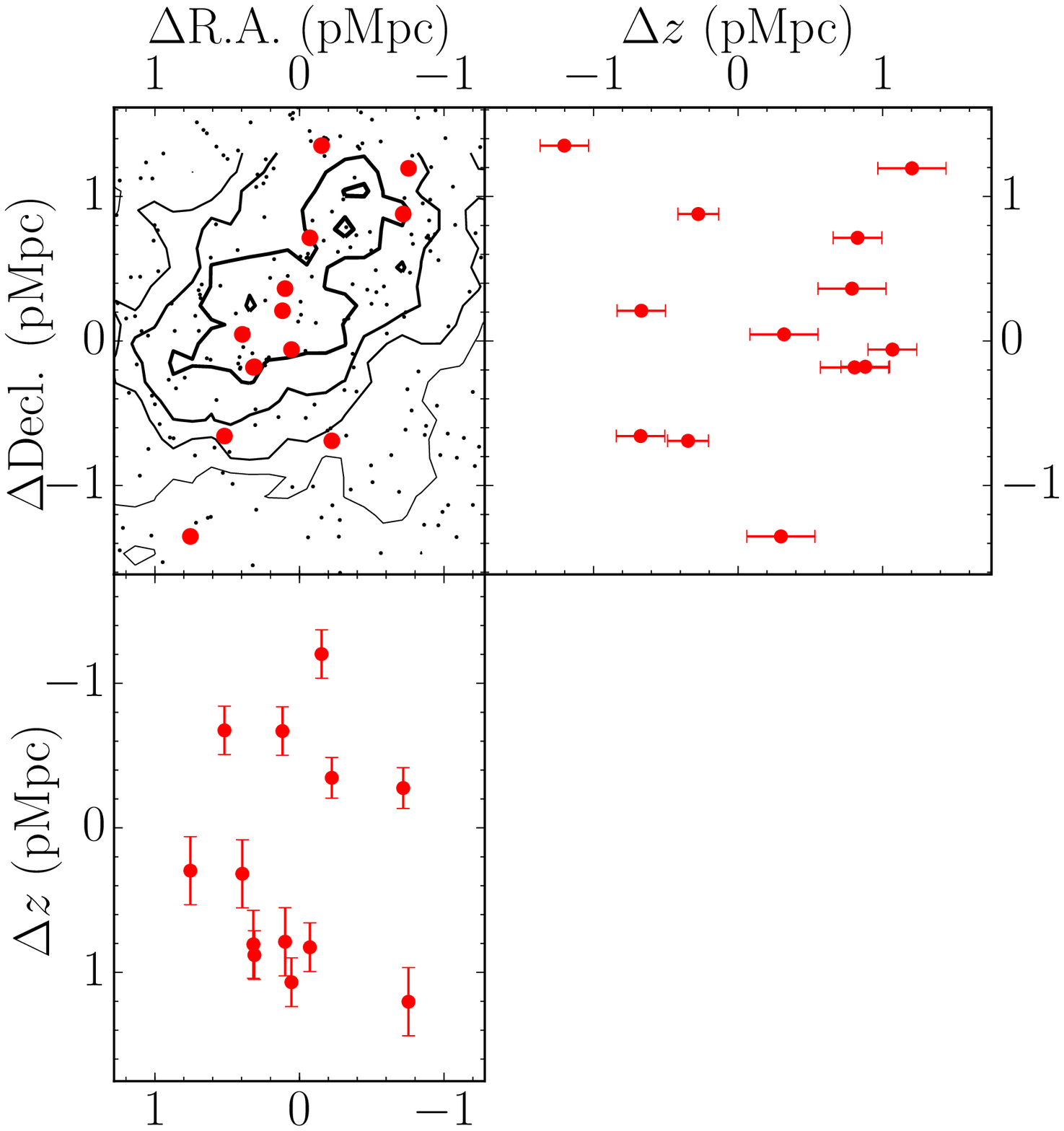}{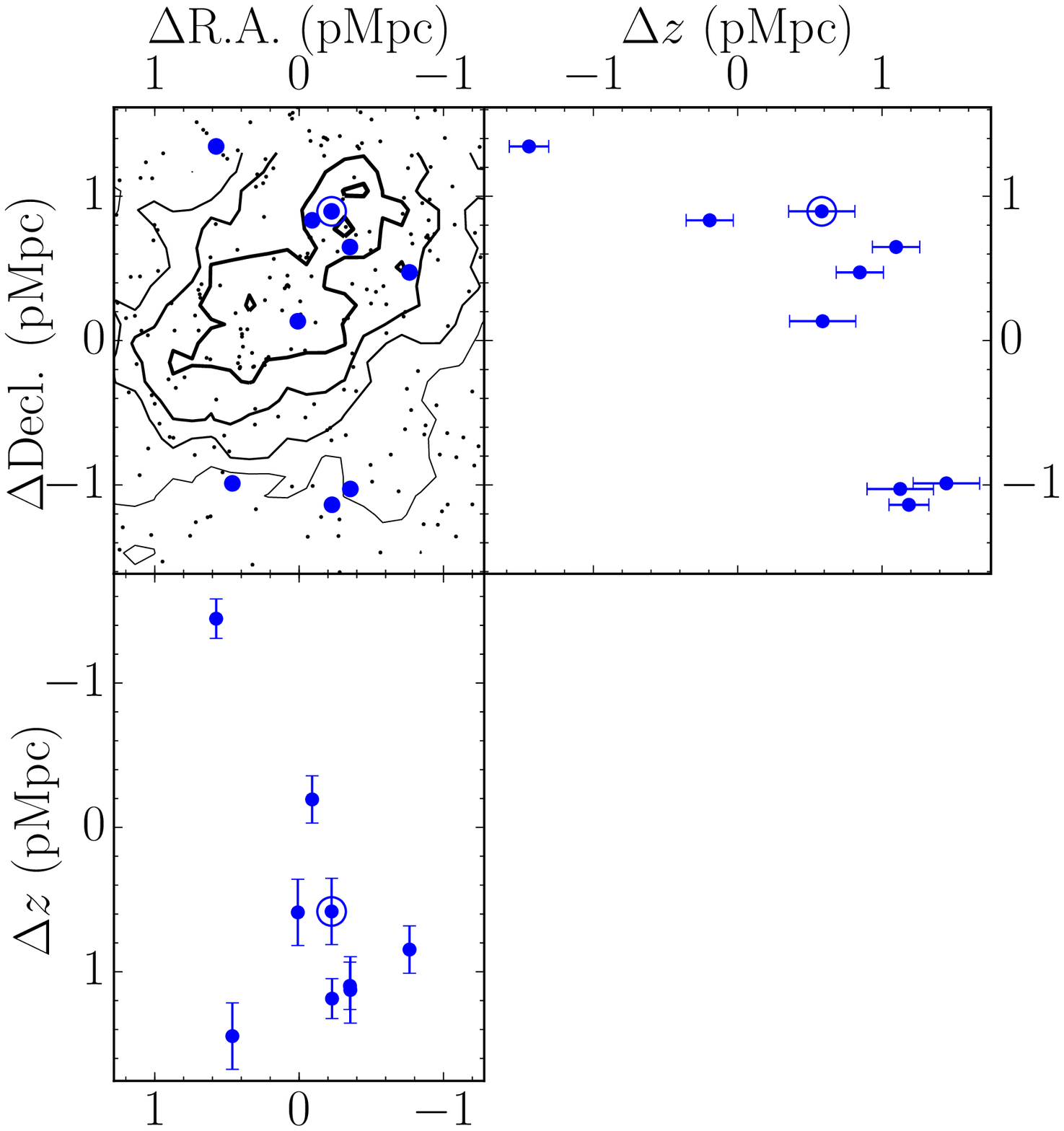}
\caption{Left-hand panel: three-dimensional distribution of the main protocluster galaxies in the D1RD01 region.
    The filled points represent the 13 protocluster galaxies, and the dots are the other $r$-dropout galaxies.
    The origin $(0,\,0)$ is $(\mathrm{R.A.},\,\mathrm{Decl.})=(22:16:50.44,\,-17:18:41.6)$.
    Right-hand panel: Same as the left-hand panel, but for the background protocluster.
    The AGN is indicated by the blue double circle.}
\label{fig:3D_gdrop}
\end{figure*}

\subsection{Internal Structure and Morphology of Protoclusters}
Next, we focus on the internal galaxy distributions of the protoclusters.
The six member galaxies of the $z=4.898$ protocluster tend to be located at the outskirt of the overdense region
rather than at the peak of overdensity ($(\Delta\mathrm{R.A.},\,\Delta\mathrm{Decl.})=(0.2,\,0.0)$).
Furthermore, three (ID=5, 6, and 17) of them seem to be strongly clustered compared with the other member galaxies.
The member galaxies of the possible protocluster at $z=3.834$ may also be distributed to avoid the center of the
protocluster since they tend to be bimodal in the redshift distribution ($p$-value of Hartigan's dip test is
0.055).
A similar internal structure was found in the protocluster at $z=6.01$ \citep{toshikawa14}, in which member
galaxies are widely distributed and divided into four subgroups.
These internal structure would be a clue to understanding the assembly process toward galaxy clusters.
In these two protoclusters, it is found that the central concentration of member galaxies is not high and small
substructures still exists.
On the other hand, it should be noted that galaxy overdensity is calculated by dropout galaxies while protocluster
members are identified by detecting Ly$\alpha$ emission.
\citet{hathi16} showed that star-forming galaxies with strong Ly$\alpha$ emission have significantly different
properties compared with those without Ly$\alpha$ emission at $z\sim2\mathrm{-}2.5$, though the absolute value of
the difference is small.
Dropout galaxies with strong Ly$\alpha$ emission are expected to have less dust, lower SFR, and less mass than
those without Ly$\alpha$ emission.
Therefore, member galaxies may be segregated in a protocluster depending on their properties: newly-formed young
galaxies may be in outskirts, and evolved massive galaxies may be near the center of a protocluster.
\citet{cooke13} also found the $\sim60\%$ of Ly$\alpha$ emitting dropout galaxies at $z\sim3$ have shell-like
spatial distribution with a radius of $\sim3\mathrm{-}6\,\mathrm{Mpc}$ in physical scale.
They concluded that dropout galaxies without Ly$\alpha$ emission tend to be in group-like environments while those
with Ly$\alpha$ emission would be distributed on the outskirts.
Their claim from the statistical method of clustering analysis could be attributed to the contribution of
protoclusters as in this study.
This implication needs to be checked by direct identification of protocluster members without Ly$\alpha$ emission
in a central region.

As for the pair of protoclusters at $z\sim3.7$, the main protocluster at $z=3.675$ is composed of at least 13
member galaxies, and nine member galaxies are confirmed for the background protocluster at $z=3.721$.
These numbers would allow us to make a close investigation of the internal structures of both protoclusters.
Figure \ref{fig:3D_gdrop} shows the three-dimensional galaxy distribution of the main and background protoclusters.
As already discussed in T16, the member galaxies of the main protocluster are spherically distributed, and nearly
half the member galaxies are concentrated into the central small region.
In this study, we have added two member galaxies, and this trend is maintained.
On the other hand, the background protocluster can be divided into three subgroups.
Each subgroup includes three (ID=21, 22, 46), five (ID=23-26, and AGN), and one (ID=47) galaxies located around 
$(\Delta\mathrm{R.A.},\,\Delta\mathrm{Decl.},\,\Delta z)=(0.0,\,-1.1,\,1.3)$, $(-0.3,\,0.6,\,0.6)$,
$(0.6,\,1.4,\,-1.5)$, respectively.
Except for the AGN, we cannot find significant differences of galaxy properties ($M_\mathrm{UV}$ and Ly$\alpha$
$EW_0$) between the main and background protoclusters; however, these two protoclusters appear to have clearly
different internal structures as shown in Figure \ref{fig:3D_gdrop}.

In addition to this visual inspection, a three-dimensional ellipsoid is fitted to the distributions of the member
galaxies following the method in \citet{lovell18}, so that we can quantitatively investigate the shape of
protoclusters.
Although an ellipsoid may be too simple modeling, it is useful to find a overall shape. 
The best-fitting ellipsoid can be determined from the eigenvalues of the moment of inertia tensor:
\begin{equation}
{\bf I}_{ij} = \sum_{n=1}^{N_\mathrm{gal}} ({\bf r}^2_n\delta_{ij} - r_{n,i}r_{n,j}),
\end{equation}
where $N_\mathrm{gal}$ is the number of member galaxies, ${\bf r}_n$ is the position vector of the $n$th galaxies,
and $i$ and $j$ are the tensor indices.
We set no weight on each member galaxy to estimate the inertia tensor.
The lengths of the primary, secondary, and tertiary axes ($a_1$, $a_2$, $a_3$) can be shown by eigenvalues,
$I_1\geq I_2\geq I_3$, as:
\begin{eqnarray}
a_1 = \sqrt{\frac{5}{2N_\mathrm{gal}}(I_1+I_2-I_3)}, \\
a_2 = \sqrt{\frac{5}{2N_\mathrm{gal}}(I_3+I_1-I_2)}, \\
a_3 = \sqrt{\frac{5}{2N_\mathrm{gal}}(I_2+I_3-I_1)}.
\end{eqnarray}
Using these axis lengths, we calculate the parameters of sphericity, $s=a_3/a_1$, and triaxiality,
$T=(a_1^2-a_2^2)/(a_1^2-a_3^2)$.
Where shape is spherical (aspherical), $s$ approaches 1 (0).
We use $T$ together to quantify the form of asphericity: oblate ($a_1>a_2,a_3$) and prolate ($a_1,a_2>a_3$)
ellipsoids have $T\sim1$ and $T\sim0$, respectively.
Table \ref{tab:shape} shows these shape parameters for the main and background protoclusters.
As expected by the visual inspection, these two protoclusters seem to have different shapes, especially in the
parameter $T$. 

\begin{figure}
\vspace{-12mm}
\epsscale{1.1}
\plotone{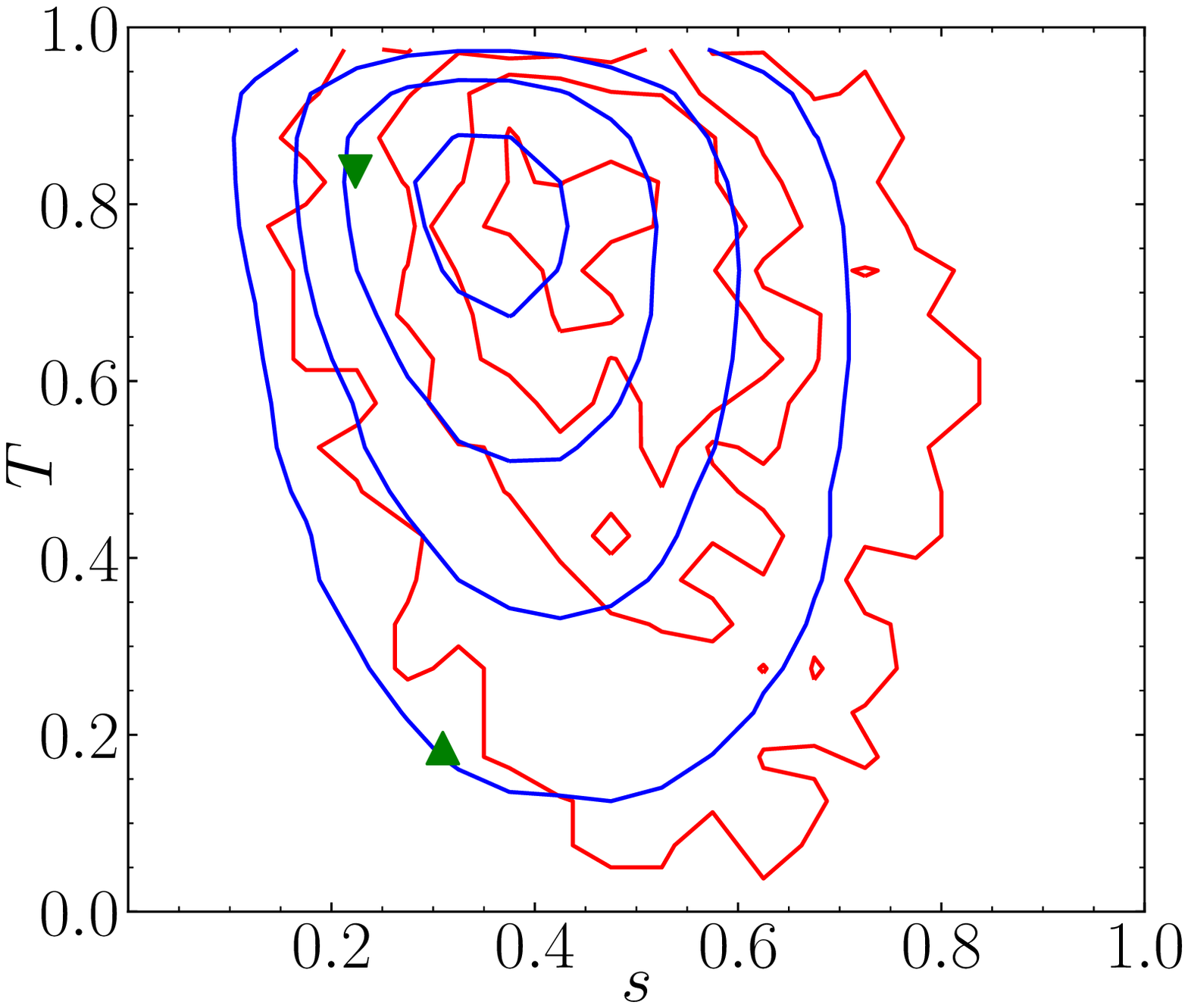}
\vspace{-4mm}
\caption{Shape parameters ($s$ and $T$) of protoclusters at $z\sim3.7$.
    The red and blue contours show the expected distribution of intrinsic and apparent values, respectively.
    The lines from inner to outer correspond to the 25\%-, 50\%-, 75\%-, and 90\%-tile contours.
    The upward and downward triangles are observed values for the main and background protoclusters, respectively.}
\label{fig:shape}
\vspace{-4mm}
\end{figure}

Our identification of protocluster members is far from complete because the fraction of spectroscopically-observed
dropout galaxies is 53\% in this overdense region, and only a portion of dropout galaxies have Ly$\alpha$ emission.
This incompleteness would strongly affect the shape estimate since it is based on only about ten galaxies.
Thus, we have deduced their intrinsic shape parameters as below.
First, we select protoclusters at $z\sim3.7$ from the Millennium simulation \citep{springel05}; the definition of
a protocluster is all halos at $z\sim3.7$ which will merge into a single halo having $>10^{14}\,\mathrm{M_\sun}$
at $z=0$.
Second, we pick up protocluster member galaxies which are similar to dropout galaxies based on galaxy properties,
such as stellar mass, SFR, or age, predicted by a semi-analytic model \citep{henriques15}.
Counterpart galaxies in the theoretical model is determined by $\mathrm{SFR}>5\,\mathrm{M_\sun\,yr^{-1}}$, which
corresponds to the limiting UV luminosity in our observation \citep{kennicutt98}.
Then, the shape parameters of $s$ and $T$ are calculated for each protocluster by using all member galaxies, which
are regarded as intrinsic parameters.
In this estimate of intrinsic parameters, there are two assumptions of descendant halo mass and SFR, which are
used to pick up protoclusters and member galaxies from the simulation, respectively.
However, the large redshift window of dropout selection could enable us to identify only more massive
protoclusters than our assumption.
The conversion from apparent (dust-uncorrected) UV luminosity to SFR involves some systematic errors due to lack
of information on dust attenuation, metallicity, or stellar age.
Therefore, we have checked possible systematic errors on the estimate of shape parameters and confirmed that shape
parameters are not significantly dependent on these changes. 
The details are shown in Appendix \ref{app:err}.
Finally, in order to account for the effect of observational bias, the same number of member galaxies as observed
ones ($N_\mathrm{gal}=13$ and 9 for the main and background protoclusters respectively) are randomly extracted
from a simulated protocluster.
In this selection of protocluster members from the theoretical model, we have also applied the spatial and
redshift windows of $L_z$ and $L_\mathrm{sky}$, which are the same as in our observation.
It should be noted that the spatial and redshift windows of our observation are large enough to impart no
significant bias to shape parameters (Appendix \ref{app:err}).
By using randomly selected member galaxies, shape parameters are re-calculated, and we investigated how different
they are from intrinsic ones.
This random realization is repeated 300 times for each protocluster.
As shown in Figure \ref{fig:shape}, random sampling results in a systematic offset on the distribution of $s$,
though the distribution of $T$ is not significantly affected.

\begin{deluxetable}{ccccc}
\tabletypesize{\normalsize}
\tablecaption{Shape parameters of the protoclusters at $z\sim3.7$ \label{tab:shape}}
\tablewidth{0pt}
\tablehead{ & \multicolumn{2}{c}{Observed} & \multicolumn{2}{c}{Expected intrinsic} \\
            & $s$ & $T$ & $s$ & $T$}
\startdata
D4GD01 & 0.31 & 0.19 & $0.45\pm0.12$ & $0.47\pm0.20$ \\
D4GD01-back & 0.22 & 0.84 & $0.37\pm0.10$ & $0.80\pm0.23$ \\
\enddata
\end{deluxetable}

Comparing the shape parameters of the observed protoclusters with the simulated ones including random sampling,
we have found that the main protocluster exhibits an unusual shape (in the 90 percentile), while the background
protocluster has a typical shape (in the 50 percentile).
Table \ref{tab:shape} also shows the expectations of intrinsic values of the shape parameters for the observed
protoclusters, which are determined by using the relation between the shape parameters calculated from full number
of galaxies and random sampling based on the theoretical model.
The significance of the shape difference between the main and background protoclusters is found to be $1.6\sigma$.
The main and background protoclusters indicate pancake-like and filamentary shapes, respectively.
Although the significance of the difference is marginal, the following is one of the possible interpretations.
As the background protocluster can be divided into three subgroups, it might be on the earlier stage of cluster
formation; thus, galaxies or small groups are just accreting along the filamentary cosmic web.
On the other hand, in the main protocluster, such building blocks of a galaxy cluster might be merging into a
single structure.

\subsection{Redshift Evolution}
\begin{figure}
\epsscale{1.2}
\plotone{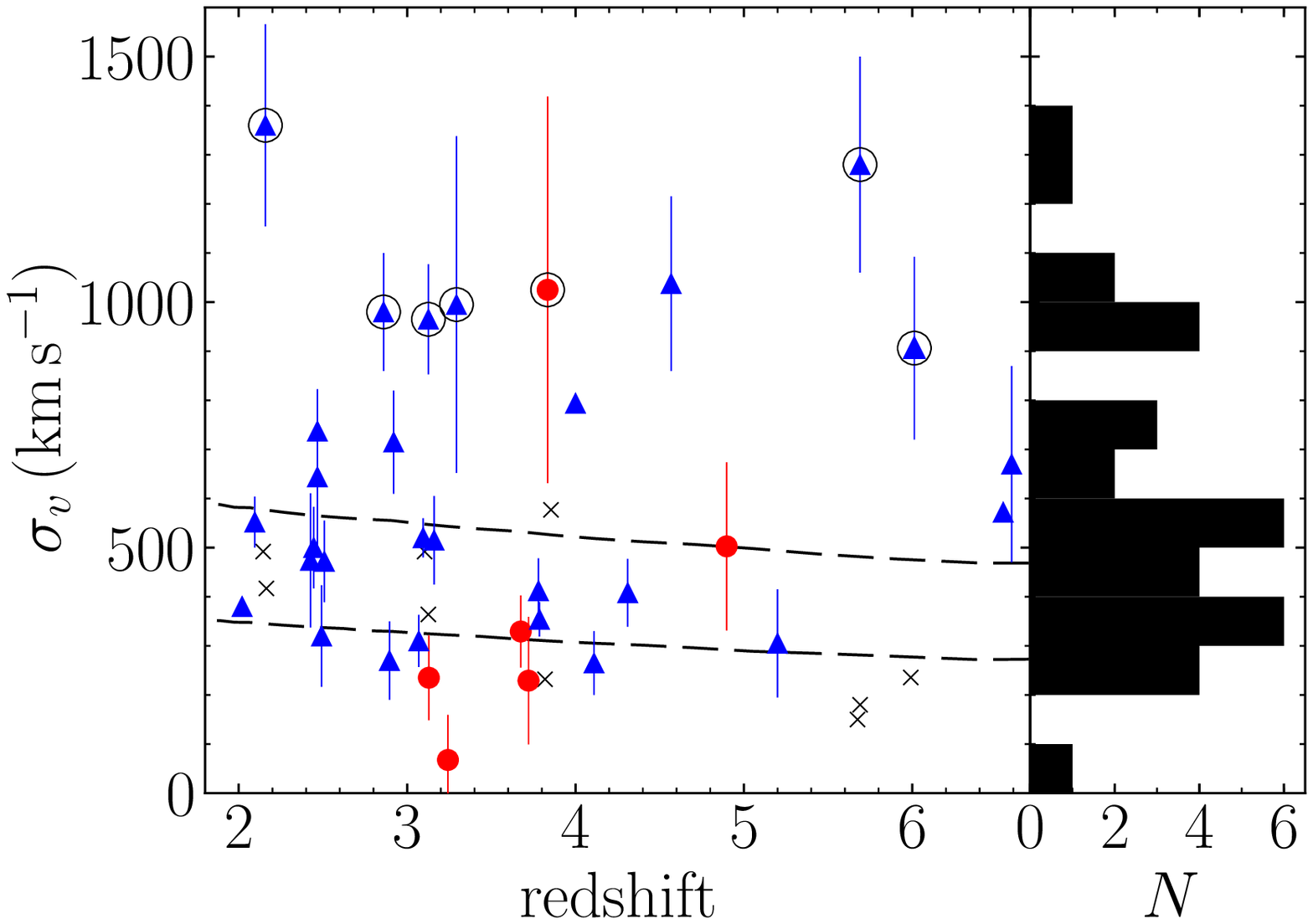}
\vspace{-9mm}
\caption{Left-hand panel: velocity dispersion of protoclusters as a function of redshift.
    The red points show protoclusters discovered by our protocluster search in the CFHTLS
    \citep[this study and][]{toshikawa16}, and the blue points are protoclusters from the literature
    \citep{chanchai19,cucciati14,cucciati18,dey16,galametz13,harikane19,kuiper11,kuiper12,lemaux14,lemaux18,miller18,ouchi05,oteo18,topping16,toshikawa14,venemans07,yuan14}.
    The sample of \citet{cucciati18} includes the protoclusters discovered by
    \citet{casey15,chiang15,diener15,wang16}.
    Protoclusters which are reported to have subgroups or deviation from normal distribution in their velocity
    distributions of member galaxies are indicated by the open black circles.
    The velocity dispersion of subgroups, if available, are represented by the black crosses.
    The dashed lines show the redshift evolution of dark matter velocity dispersion of
    $1\times10^{14}\,\mathrm{M_\sun}$ and $5\times10^{14}\,\mathrm{M_\sun}$ halos at $z=0$.
    The redshift evolution of velocity dispersion is derived from that of dark matter halo mass by assuming virial
    equilibrium and extended Press-Schechter model.
    Right-hand panel: histogram of the velocity dispersion of protoclusters.
    The velocity dispersion of subgroups is not included in this histogram.
    It should be noted that \citet{wang16} detected extended X-ray emission; thus, it should be categorized as
    cluster not protocluster.
    However, since cluster formation would be seamless from high to low redshifts, the cluster found by
    \citet{wang16} is also plotted on this figure as a example at $z=2.5$.}
\label{fig:vel-dis}
\vspace{-2mm}
\end{figure}

We have characterized the large-scale and internal structures of the protoclusters, which may help us understand
cluster formation in the context of the hierarchical structure formation by combining with other protoclusters
from the literature.
It should be noted that protocluster samples searched by various methods may be heterogeneous, and the definitions
of protoclusters differ depending on studies.
The sizes of some protoclusters are also artificially limited by the size of the FoV of their associated
observations.
In this study, we use velocity dispersion, which indicates the dynamical state of protoclusters, to compare with
other protoclusters.
In the estimate of the velocity dispersion, we assume that the redshift difference of protocluster members is
attributed to the velocity difference, instead of the difference of line-of-sight distance.
The biweight scale \citep{beers90} is used as the estimator of velocity dispersion (Table \ref{tab:pcl}).
In addition, we have compiled the velocity dispersion of known protoclusters at $z>2$ from the literature and
investigated the relation between redshift and velocity dispersion (Figure \ref{fig:vel-dis}).
We cannot find a significant correlation though velocity dispersion is expected to be increasing with protocluster
growth.
The heterogeneous sample of protoclusters from the collection of many previous studies might dilute a possible
trend between velocity dispersion and redshift because the relation is also dependent on the descendant halo mass
at $z=0$.
However, we can find a distinguishing feature in the histogram of velocity dispersion (the right panel of Figure
\ref{fig:vel-dis}).
While most protoclusters have velocity dispersions of $\sim200\mathrm{-}600\,\mathrm{km\,s^{-1}}$, some have large
velocity dispersions of $\sim1000\,\mathrm{km\,s^{-1}}$.
This cause the skewed distribution of velocity dispersion as shown in the right panel of Figure \ref{fig:vel-dis}.
The skewness of this distribution is found to be 0.71, and the null hypothesis that the observed distribution is
generated from a single normal distribution can be rejected with 97\% significance level according to the
Anderson-Daring test.
The lower peak is almost consistent with the redshift evolution of dark mater velocity dispersion of typical galaxy
clusters ($\sim1\mathrm{-}5\times10^{14}\,\mathrm{M_\sun}$); on the other hand, it would be difficult to explain
the higher peak by halo evolution under virial equilibrium.
Either the protoclusters having higher velocity dispersion ($\gtrsim1000\,\mathrm{km\,s^{-1}}$)
could contain subgroups or the velocity (redshift) distribution of protocluster members may deviate from normal
distribution
\citep[the circles in Figure \ref{fig:vel-dis}; e.g.,][]{kuiper12,lemaux14,toshikawa14}.
Some previous studies measured the velocity dispersion of each individual subgroup (the crosses in Figure
\ref{fig:vel-dis}), and they are almost comparable to that of the lower peak.

In the local universe, galaxy clusters of $\sim3\times10^{14}\,\mathrm{M_\sun}$ halos have velocity dispersion of
$\sim500\,\mathrm{km\,s^{-1}}$, and, even for massive clusters of $\sim1\times10^{15}\,\mathrm{M_\sun}$ halos,
their velocity dispersion is $\lesssim1000\,\mathrm{km\,s^{-1}}$ \citep[e.g.,][]{berlind06,tempel14,wilson16}.
On the other hand, merging galaxy clusters, which can be traced by using radio relics, are found to show high
velocity dispersions \citep[$\gtrsim1000\,\mathrm{km\,s^{-1}}$:][]{golovich17}.
Therefore, higher velocity dispersion can be explained if such protoclusters are in a merging phase of galaxy
groups on the way to forming more massive structures; thus their dynamical state may be far from virialization.
Systematic merger motions would need to be included in the calculation of velocity dispersion in addition to the
random motion of protocluster members.
\citet{kuiper11} have simulated the evolution of velocity dispersion in the case of the Spiderweb protocluster.
They found that velocity dispersion is dramatically changed from $\sim400\mathrm{-}500\,\mathrm{km\,s^{-1}}$ to
$\sim900\,\mathrm{km\,s^{-1}}$ at the point of merger of two halos.
As shown in Figure \ref{fig:vel-dis}, we find that protoclusters with higher velocity dispersion account for about
one forth of the total, distributed over whole redshift range.
Protoclusters would be evolving by mergers of galaxy groups as well as steady galaxy accretion, and these two
phases are repeated from early to late developmental stages.
It should be noted that there is a possibility that higher velocity dispersion has been overestimated by
classifying independent groups as a single protocluster.
As discovered by \citet{cucciati18}, protoclusters would bring together many smaller components and form
large-scale structures.
It is necessary to map out galaxy distribution beyond the scale of a protocluster in order to discuss how they
grow into mature clusters.

\section{Conclusion \label{sec:conc}}
In this study, we have presented optical follow-up spectroscopy on the three overdense regions of $g$- and
$r$-dropout galaxies in the CFHTLS Deep Fields.
In the overdense region of D4GD01, the existence of a protocluster is already confirmed at $z=3.675$ by
\citet{toshikawa16}.
This study increases the number of confirmed member galaxies of this protocluster to thirteen.
In addition, we have newly discovered a protocluster including nine member galaxies at $z=3.721$ at the same sky
position with the $z=3.675$ protocluster.
As for the overdense region of D1RD01, we have confirmed a protocluster at $z=4.898$, composed of six member
galaxies.
Furthermore, a possible protocluster is found at $z=3.834$ in the overdense region of D1GD02, though it is a
tentative detection.
From these protoclusters, including a possible one, we have obtained the major implications as below.

In the vicinity of the $z=4.898$ protocluster, there are two small galaxy groups, each including three galaxies.
Since the separations between these two groups and the $z=4.898$ protocluster are only $\Delta z\sim0.05$, these
two groups are expected to become a part of a supercluster at $z=0$, rather than merge into the protocluster to
form a single massive dark matter halo.
Similarly, in the overdense regions of both D1GD02 and D4GD01, we have found close pair-like structures, whose
redshift separation is only $\Delta z\sim0.05$. 
These results suggest that large-scale galaxy/group assembly comparable to the size of superclusters start by
$z\gtrsim4$, and the primordial satellite components of superclusters appear at $z\sim4\mathrm{-}5$ in parallel
with the formation of central protoclusters.
It should be noted that this conclusion depends on descendant halo mass; if this protocluster is the progenitor of
a significantly rich cluster ($>10^{15}\,\mathrm{M_\sun}$), it is possible to incorporate the neighboring groups
into a single halo by $z=0$.

For the protocluster pair at $z\sim3.7$ in the D4GD01 overdensity, their detailed internal structures are
investigated by fitting a triaxial ellipsoid to the distribution of member galaxies.
In this analysis, after carefully considering sampling bias based on theoretical models, we have tentatively found
that the two protoclusters have different shapes ($1.6\sigma$ significance).
The main protocluster, which has thirteen member galaxies, has a pancake-like shape, while the other protocluster,
which is located just behind the main protocluster, and composed of nine galaxies, exhibits a filamentary shape.
The background protocluster can be divided into three subgroups.
These three groups nearly align in three-dimensional space as suggested by the ellipsoid fitting.
This probably indicates that they are on the way to merging along cosmic web to make a single dark matter halo.
On the other hand, the main protocluster would be expected to develop more than the background protocluster,
judging from the number of confirmed member galaxies and its shape.

We have also discussed the redshift evolution of protoclusters based on their velocity dispersion by combining
with other known protoclusters from the literature.
Although there is no significant dependence of protocluster velocity dispersion on redshift, we have found a
distribution skewed towards high velocity dispersion.
This could be interpreted as the two phases of cluster formation, which are steady galaxy accretion and mergers of
galaxy groups.
This would be generally consistent with the picture of hierarchical structure formation model.
Although it is difficult to perform quantitative investigations due to the small and heterogeneous sample of
protoclusters, the incidence of mergers or mass ratio between merging groups will be keys to understand the
formation of galaxy clusters.

As we have shown, the protoclusters are characterized from the viewpoints of shape and large-scale structure.
We have found that the formation of a supercluster starts in the early universe, and the main and background
protoclusters at $z\sim3.7$ show different galaxy distributions.
However, since our results are derived from a few protoclusters based on the investigation of a single galaxy
population, it is difficult to evaluate whether they are representative of all protoclusters or not.
Through the multi-wavelength observations of more protoclusters, we will be able to reveal how the large-scale
structure is built from the early to present-day universe, which is related to the cosmological parameters and the
initial perturbations of mass density.
Furthermore, this is linked to galaxy evolution across cosmic time because environments have and important
influence on star-formation activity.
In the future, we will systematically observe protoclusters provided by the Hyper SuprimeCam Subaru Strategic
Program \citep{toshikawa18} in order to discuss the dynamical evolution of protoclusters and its relation with the
physical properties of member galaxies.

\acknowledgments
We are really grateful to Dr. Brian Lemaux for valuable comments and suggestions that significantly improved the
manuscript.
The CFHTLS data used in this study are based on observations obtained with MegaPrime/MegaCam, a joint project of
CFHT and CEA/IRFU, at the Canada-France-Hawaii Telescope (CFHT), which is operated by the National Research
Council (NRC) of Canada, the Institut National des Science de l'Univers of the Centre National de la Recherche
Scientifique (CNRS) of France, and the University of Hawaii.
This study is based in part on data products produced at Terapix available at the Canadian Astronomy Data Centre
as part of the Canada-France-Hawaii Telescope Legacy Survey, a collaborative project of NRC and CNRS.
This study is also based on data collected at the W. M. Keck telescope, which is operated as a scientific
partnership among the California Institute of Technology, the University of California, and the National
Aeronautics and Space Administration.
The W. M. Keck Observatory was made possible by the generous financial support of the W.M. Keck Foundation.
We are grateful to the W. M. Keck Observatory staff for their help with the observations, and we wish to
recognize and acknowledge the very significant cultural role and reverence that the summit of Mauna Kea has
always had within the indigenous Hawaiian community.
This research was supported by the Japan Society for the Promotion of Science through Grants-in-Aid for Scientific
Research 18K13575 and 18J01430.
NK acknowledges support from the JSPS grant 15H03645, and RAO is grateful for financial support from FAPERJ,
CNPq, and FAPESP.

\facilities{CFHT (MegaCam), Keck:II (DEIMOS)}

\appendix
\section{Systematic error on shape parameters \label{app:err}}
We make some assumptions to predict the intrinsic shape parameters from observed protoclusters based on the
theoretical model.
These are mainly the descendant halo mass, SFR, and observing window size.
In the following subsections, we have evaluated how large these assumptions make an effect on the estimate of
shape parameters.

\subsection{Descendant Halo Mass}
Ideally, protoclusters are defined as the progenitors of $>10^{14}\,\mathrm{M_\sun}$ halos at $z=0$.
However, the large redshift window of dropout selection would dilute the signal of protoclusters, which causes 
lower completeness for the progenitors of lower-mass clusters.
The average of $>10^{14}\,\mathrm{M_\sun}$ halos at $z=0$ is $\sim2\times10^{14}\,\mathrm{M_\sun}$ while the
expected average descendant halo mass from the clustering analysis or abundance matching is found to be
$\sim5\times10^{14}\,\mathrm{M_\sun}$ \citep{toshikawa18}.
When the assumption of halo mass limit is changed to $>3\times10^{14}\,\mathrm{M_\odot}$, the average halo mass
turns to $\sim5\times10^{14}\,\mathrm{M_\odot}$.
Thus, we have also calculated the shape parameters of protoclusters which are the progenitors of
$>3\times10^{14}\,\mathrm{M_\odot}$ halos at $z=0$.
The left panel of Figure \ref{fig:app_err} shows the difference of shape parameters depending on descendant halo
mass at $z=0$.
Although the progenitors of higher-mass clusters tend to be more spherical (higher $s$), the significance of
difference is small ($0.4\sigma$).

\subsection{Star Formation Rate}
Observable protocluster members in simulated protoclusters are assumed to be
$\mathrm{SFR}>5\,\mathrm{M_\sun\,yr^{-1}}$.
Although this criterion should be determined by considering both UV luminosity and dust attenuation of our dropout
galaxies, it is difficult to correctly estimate dust attenuation based on our dataset.
Thus, we have tested on the cases of $\mathrm{SFR}>2.0$ and $>20.0\,\mathrm{M_\odot\,yr^{-1}}$ in order to evaluate
the effect of the uncertainty of dust attenuation, and the difference is found to be $0.5\sigma$ significance (the
middle panel of Figure \ref{fig:app_err}).

\subsection{Observing Window Size}
\begin{figure}
\epsscale{1.1}
\plotone{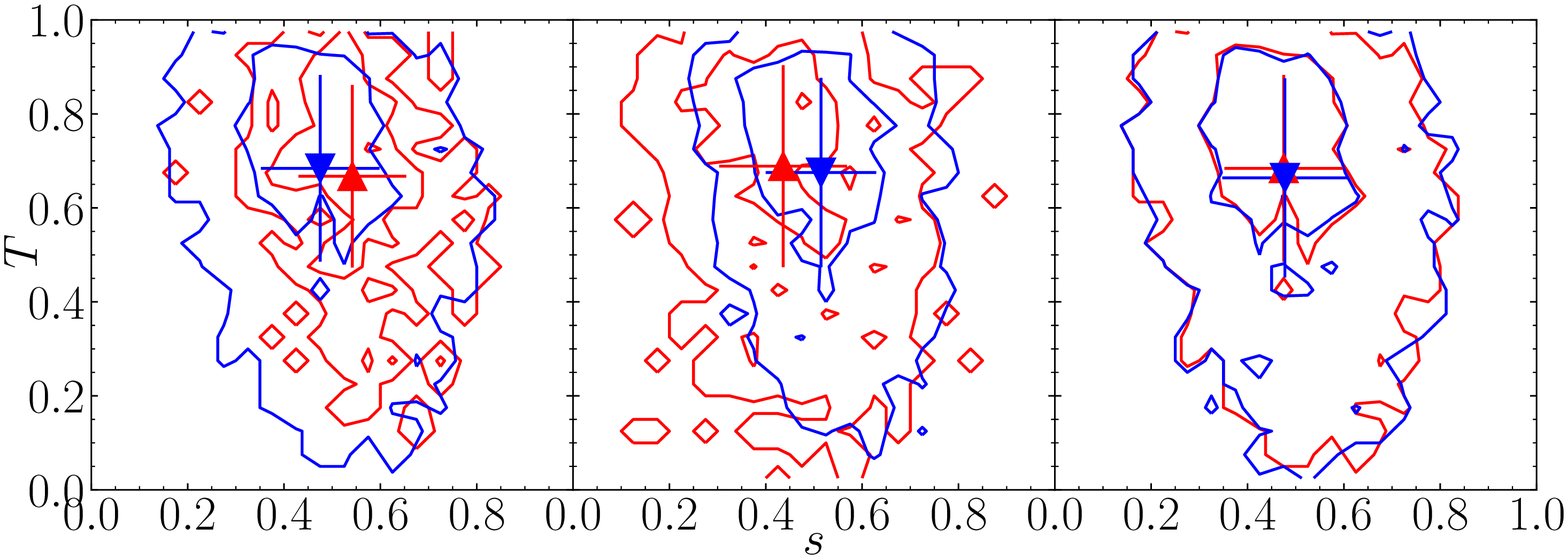}
\caption{Comparison of the distribution of shape parameters under various assumptions.
    Left: The blue and red contours show the shape parameters of the progenitors of
    $>1\times10^{14}\,\mathrm{M_\odot}$ and $>3\times10^{14}\,\mathrm{M_\odot}$ halos, respectively.
    The inner and outer lines indicate the 50\%- and 95\%-tile, respectively.
    The red contours are noisier than the blue contours due to smaller sample size caused by its higher mass limit.
    The triangles with error bars are the average and $1\sigma$ uncertainty for the two cases (the color code is
    the same as the contour lines).
    Middle: The blue and red contours show the shape parameters estimated by using protocluster members with
    $\mathrm{SFR}>2.0$ and $>20.0\,\mathrm{M_\odot\,yr^{-1}}$, respectively.
    The lines and points have the same meaning as the left panel.
    Right: The blue and red show the shape parameters estimated by protocluster members within the observing
    window of our follow-up spectroscopy and all ones, respectively.}
\label{fig:app_err}
\end{figure}

If observing window is smaller than the size of protoclusters, it could be difficult to properly estimate shape
parameters.
Protocluster members are typically spread over $\sim2\,\mathrm{arcmin}$ radius, or $\sim5\,\mathrm{arcmin}$ at
maximum.
Our follow-up spectroscopy is performed with Subaru/FOCAS and Keck/DEIMOS, whose FoVs are a circle with
$3\,\mathrm{arcmin}$ radius and a rectangle with $16\,\mathrm{arcmin}\,\times\,4\,\mathrm{arcmin}$ respectively.
Although these FoVs are larger than typical size of the distribution of protocluster members, some surrounding
protocluster members are outside the observing area.
Those surrounding members may have a large impact on the estimate of shape parameters.
We have checked how our observing window alters shape parameters.
As shown in the right panel of Figure \ref{fig:app_err}, the results are consistent with each other, suggesting
our observing window is found to have a sufficient size.

{}

\end{document}